\documentclass[aps,prd,reprint,preprintnumbers,nofootinbib]{revtex4-1}
\pdfoutput=1

\usepackage{amsmath}
\usepackage{amsfonts}
\usepackage{amssymb}
\usepackage{graphicx}
\usepackage{bbold}
\usepackage{mathtools}
\usepackage{multirow}
\usepackage{enumitem} 
\usepackage{hyperref}
\allowdisplaybreaks
\usepackage[dvipsnames]{xcolor}

\allowdisplaybreaks
\newcommand{\ket}[1]{|#1\rangle}

\newcommand{\orcid}[1]{\href{https://orcid.org/#1}{#1}}

\begin{document}

\title{Compact perturbative expressions for oscillations with sterile neutrinos in matter}

\author{Stephen J.~Parke}
\email{parke@fnal.gov}
\thanks{\orcid{0000-0003-2028-6782}}
\affiliation{Theoretical Physics Department, Fermi National Accelerator Laboratory, Batavia, IL 60510, USA}

\author{Xining Zhang}
\email{xining@uchicago.edu}
\thanks{\orcid{0000-0001-8959-8405}}
\affiliation{Enrico Fermi Institute and Department of Physics, University of Chicago, Chicago, Illinois 60637, USA}

\preprint{FERMILAB-PUB-19-042-T}

\date{\today}

\begin{abstract}
We extend a simple and compact method for calculating the three flavor neutrino oscillation probabilities in uniform matter density to schemes with sterile neutrinos, with favorable features inherited. The only constraint of the extended method is that the scale of the matter potential is not significantly larger than  the atmospheric $\Delta m^2$, which is satisfied by all the running and proposed accelerator oscillation experiments. Degeneracies of the zeroth order eigensystem around solar and atmospheric resonances are resolved. Corrections to the zeroth order results are restricted to no larger than the ratio of the solar to the atmospheric $\Delta m^2$. The zeroth order expressions are exact in vacuum because all the higher order corrections vanish when the matter potential equals zero. Also because all the corrections are continuous functions of matter potential, the zeroth order precision is much better than $\Delta m^2_\odot/\Delta m^2_\text{atm}$ for weak matter effect. Numerical tests are presented to verify the theoretical predictions of the exceptional features. Precision and speed comparisons with previous 3+1 methods are performed. Moreover, possible applications of the method in experiments to check the existence of sterile neutrinos are discussed.   
\end{abstract}

\maketitle

\section{Introduction}
Since the discovery of neutrino oscillations, \cite{Fukuda:1998mi},  which determined that neutrinos are massive particles, many studies of neutrino scenarios beyond the  three-flavor Standard Model have been performed.

One promising solution to the origin of the neutrino masses is a theoretical scheme with additional sterile neutrinos. In such a scheme neutrino oscillations will be modified because of the additional mixing with sterile neutrinos. In matter, calculations of neutrino propagation will be significantly more complicated since the sterile neutrinos also change the Wolfenstein matter effect term \cite{Wolfenstein:1977ue} in the Hamiltonian. There have been some analytical derivations of the matter effect in a 3+1 scenario, i.e. one sterile neutrino  \cite{Li:2018ezt} in addition to the 3 active ones. However, the exact analytical solutions are impossible for more than one sterile neutrino because a quintic or even higher order equation will be encountered. Consequently, alternative perturbation approaches should be considered.

A satisfying perturbative framework, regardless of the existence of sterile neutrinos, is expected to possess the following properties: the expansion parameter is small; crossings of zeroth order eigenvalues are avoided anywhere; the approximated values go to the exact ones in vacuum. Recently, a compact perturbative framework achieving all the objectives above was developed by Denton, Minakata and Parke (DMP) to calculate the propagation of neutrinos in matter under the assumption of the standard three-flavor scheme \cite{Minakata:2015gra,Denton:2016wmg,Denton:2018fex}.  

The main focus in this paper is to extend the principle and method of the DMP framework to schemes with sterile neutrinos when the scale of matter potential $a$ is smaller or comparable to $\Delta m^2_\text{atm}$, which is the case of all running and proposed accelerator neutrino oscillation experiments. The expansion parameter \cite{Denton:2016wmg,Nunokawa:2005nx,Parke:2016joa}, which will be retained by the extension, is
\begin{align}
    \epsilon&\equiv\Delta m^2_{21}/\Delta m^2_{ee}\simeq 0.03,  \notag \\
\Delta m^2_{ee}&\equiv \cos^2\theta_{12}\Delta m^2_{31}+\sin^2\theta_{12}\Delta m^2_{32}. \label{eq:epsilon}
\end{align}
The perturbative Hamiltonian will have no diagonal elements, and all its off-diagonal elements are proportional to $\epsilon$ and vanish in vacuum. Crossings of the zeroth order active eigenvalues will be resolved by a series of real or complex rotations; whereas crossings of the large sterile eigenvalues  will not be considered since this will happen only if the matter effect is extremely large. 

The structure of this paper is listed as follows: in Section \ref{sec:rotations}, we derive details of the rotations. This gives the zeroth order PMNS matrix and eigenvalues. The perturbative Hamiltonian is also determined by the rotations. In Section \ref{sec:perturbation}, we discuss the higher order corrections by perturbative expansions after the rotations. A numerical test will also be presented to verify the predicted precision. In Section \ref{sec:probability} we use these perturbative expressions to calculate the oscillation probabilities of different channels and baselines. Moreover, potential applications of the method are discussed. We compare our method to some former works in Section \ref{sec:compare}. Section \ref{sec:conclusion} is the conclusion. All other remarks and
supplementary materials that are useful can be found in the Appendices.

\section{Rotations to derive zeroth order approximations and perturbative Hamiltonian}\label{sec:rotations}

The principle of the method in \cite{Minakata:2015gra,Denton:2016wmg,Denton:2018fex} is that by implementing a series of rotations of the Hamiltonian, one can disentangle the crossings of the diagonal elements and diminish the off-diagonal elements to arbitrary scales. In particular:    
\begin{enumerate}
\item In the given Hamiltonian in flavor basis, find the sector with leading order (largest absolute value) off-diagonal element, then perform a rotation to diagonalize this sector. 
\item Use the rotated Hamiltonian to replace the initial one and repeat the process until all the off-diagonal elements are smaller than the expected scale and the diagonal element crossings are eliminated.   
\end{enumerate} 
In principle, the above process is not designated to any specific dynamical system and is also applicable to the schemes with sterile neutrinos.

However, this scheme must be implemented with considerable care otherwise the resulting analytic expressions becoming extremely  long and complicated.   First, one has to carefully choose the extension to the PMNS matrix to include sterile neutrinos as the standard choice here is far from optimal.  Second, one has to decide whether or not one deals  with all level crossings of the diagonal elements of the Hamiltonian or restrict the range of applicability of the result.  We address these issues in depth in the following sub-sections.

\subsection{PMNS matrix in vacuum} \label{subsec:convention}
If we assume a $3+N$ scheme, i.e. there are $N$ sterile neutrinos in the scheme, the Hamiltonian in the flavor basis will be
\begin{widetext}
\begin{equation}
\textbf{H}=\frac{1}{2E}\Big[\textbf{U}_\text{PMNS}\,\text{diag}(0,\,\Delta m^2_{21},\,\Delta m^2_{31},\,\Delta m^2_{41},\,...,\,\Delta m^2_{N1})\,\textbf{U}^\dagger_\text{PMNS}+\text{diag}(a(x),\,0,\,0,\,b(x),\,...,\,b(x))\Big], \label{eq:Hflavor}
\end{equation}
\end{widetext}
where $a$ and $b$ are Wolfenstein's matter potentials \cite{Wolfenstein:1977ue}:
\begin{align}
a&=2\sqrt{2}G_FN_eE\simeq 1.52\times 10^{-4}\bigg(\frac{Y_e\,\rho}{\text{g}\cdot \text{cm}^{-3} }\bigg)\bigg(\frac{E}{\text{GeV}}\bigg)\text{eV}^2, \notag \\
b&=\sqrt{2}G_FN_nE. \label{eq:ab}
\end{align}
For earth matter the  neutron number density $N_n$ is approximately equal to the electron number density $N_e$ so  that $b\approx a/2$.

The PMNS matrix $\textbf{U}_\text{PMNS}$ in vacuum, which relates the flavor basis  and the mass basis, is the product of a series of (complex) rotations \cite{Maki:1962mu,Pontecorvo:1967fh}. In the Standard Model, the convention is chosen to be $\textbf{U}^\text{SM}_\text{PMNS}\equiv \textbf{U}_{23}\,\textbf{U}_\text{13}\,\textbf{U}_{12}$. In the $3+N$ scheme there will be extra rotations mixing with sterile neutrinos. It is natural to require that the convention is equivalent to that of the $3\nu$SM case if all the extra rotations are trivial. Therefore, we will keep the relative positions of the three rotation matrices in the active sector when defining the PMNS matrix with sterile neutrinos. 

Also it is observed that both the second and the third row vanish in the matter potential term in Eq.\,\ref{eq:Hflavor}, thus we will keep $\textbf{U}_{23}$ as the first rotation in the PMNS matrix so the R.H.S of Eq.\,\ref{eq:Hflavor} will be independent of the 2-3 mixing parameters if we perform the $\textbf{U}_{23}$ rotation. The last step to determine the convention of the PMNS matrix is finding places after the $\textbf{U}_{23}$ for the rotations mixing with the sterile neutrinos. By trying different choices to simplify the calculation processes, we adopt the following convention of the PMNS matrix:
\begin{equation}
\textbf{U}_\text{PMNS}\equiv \textbf{U}_{23}(\theta_{23},\delta_{23})\,\textbf{U}_\text{sterile}\,\textbf{U}_\text{13}(\theta_{13})\,\textbf{U}_{12}(\theta_{12}),\label{eq:convention}
\end{equation}
where $\textbf{U}_\text{sterile}$ is the product of all the rotations mixing with sterile neutrinos. This choice leads to significant reductions in the complexity of the calculations and the resulting expressions. Physics, of course, is independent of this choice.

In the following sections, we will use the 3+1 scheme as an example to develop the expressions for the schemes with sterile neutrinos. 
 In particular, we choose \footnote{Convention of the CP phases is chosen to simplify the calculation process. Different conventions can be related by pure phase transformations. }
\begin{equation}
\textbf{U}^\text{3+1}_\text{sterile}\equiv \textbf{U}_{34}(\theta_{34},\delta_{34})\,\textbf{U}_{24}(\theta_{24},\delta_{24})\,\textbf{U}_{14}(\theta_{14}). \label{eq:Usterile}
\end{equation}

Current global fits \cite{Dentler:2018sju,Dev:2019pdg,Diaz:2017mfd} suggest $|U_{i4}|\sim 0.1$, so in this paper we assume that $\textbf{U}_\text{sterile}\simeq\mathbb{1}+\mathcal{O}(\sqrt{\epsilon})$, which means that $s_{i4}\sim\mathcal{O}(\sqrt{\epsilon})$ for $i=1,\,2,\,3$. The small parameter $\epsilon$ is defined in Eq.\,\ref{eq:epsilon}.

The convention in Eq.\,\ref{eq:convention} is different from the usual one used by many papers in which $\textbf{U}_\text{sterile}$ comes before (i.e. on the left side of) all the three rotations in the active sector (see e.g., \cite{Blennow:2018hto}). We will derive the relations of the mixing angles and phases connecting both conventions in Appendix \ref{appendix:convention}.

\subsection{$\textbf{U}_{23}$ and $\textbf{U}_\text{sterile}$ rotations} 

We first define a rotated basis $\ket{\tilde{\nu}}$ by
\begin{align}
\ket{\tilde{\nu}}&\equiv \textbf{U}^\dagger_\text{sterile}\,\textbf{U}^\dagger_{23}\ket{\nu}_\text{f}\notag \\
=&\textbf{U}^\dagger_{14}(\theta_{14})\textbf{U}^\dagger_{24}(\theta_{24},\delta_{24})\textbf{U}^\dagger_{34}(\theta_{34},\delta_{34})\textbf{U}^\dagger_{23}(\theta_{23},\delta_{23})\ket{\nu}_\text{f},
\end{align}
$\ket{\nu}_\text{f}$ is the flavor basis. After the rotations, the Hamiltonian becomes 
\begin{align}
\tilde{\textbf{H}}&\equiv \textbf{U}^\dagger_\text{sterile}\,\textbf{U}^\dagger_{23}(\theta_{23},\delta_{23})\,\textbf{H}\,\textbf{U}_{23}(\theta_{23},\delta_{23})\,\textbf{U}_\text{sterile} \notag \\
\notag \\
&=\left(\begin{array}{cc}
\tilde{H} &  \\
 & \frac{M^2}{2E}
\end{array}\right)+\tilde{\textbf{H}}_M. \label{eq:Htilde}
\end{align}
In the above equation $M^2(b)\equiv\Delta m^2_{41}+ b\,c^2_{14}c^2_{24}c^2_{34}$, $\tilde{H}$ is a $3\times 3$ submatrix in the active sector and in $\tilde{\textbf{H}}_M$ all the elements not in the 4th column or row vanish. 

Based on the scales we can distribute the elements of $\tilde{H}$ into two parts, i.e. 
\begin{equation}
\tilde{H}=\tilde{H}_0+\tilde{H}_1.
\end{equation}
The leading order term is
\begin{widetext}
\begin{equation}
\tilde{H}_{0}=\frac{1}{2E}\left(\begin{array}{ccc}
\lambda_a & & s_{13}c_{13}\Delta m^2_{ee}+\epsilon\,bk_{13}c_{24}c_{34}e^{-i\delta_{34}}\\
 &\lambda_b & \\
 s_{13}c_{13}\Delta m^2_{ee}+\epsilon\,bk_{13}c_{24}c_{34}e^{i\delta_{34}}& &\lambda_c
\end{array}\right),
\end{equation} 
\end{widetext}
where 
\begin{equation}
k_{ij}\equiv \frac{s_{i4}s_{j4}}{\epsilon}\sim \mathcal{O}(1),\quad i,j\in\{1,2,3\}
\end{equation}
and the diagonal elements, which can be approximations to the eigenvalues are
\begin{align}
\lambda_a&=(s^2_{13}+\epsilon\,s^2_{12})\,\Delta m^2_{ee}+a\,c^2_{14}+\epsilon\,b\,k_{11}\,c^2_{24}c^2_{34}, \notag \\
\lambda_b&=\epsilon\,(c^2_{12}\,\Delta m^2_{ee}+b\,k_{22}\,c^2_{34}), \notag \\
\lambda_c&=(\,c^2_{13}+\epsilon\,s^2_{12}\,)\,\Delta m^2_{ee}+\epsilon\,b\,k_{33}. 
\end{align}
In the first order term $\tilde{H}_1$, all the diagonal elements vanish, and the off-diagonal elements are
\begin{align}
(\tilde{H}_1)_{12}&=\,\frac{\epsilon}{2E}\,\Big(c_{12}s_{12}c_{13}\,\Delta m^2_{ee}+b\,k_{12}\,c_{24}c^2_{34}\,e^{-i\delta_{34}}\Big),\notag \\
(\tilde{H}_1)_{23}&=\,\frac{\epsilon}{2E}\,\Big[-c_{12}s_{12}s_{13}\,\Delta m^2_{ee}+b\,k_{23}\,c_{34}\,e^{i(\delta_{24}-\delta_{34})}\Big], \notag \\
(\tilde{H}_1)_{13}&=0.
\end{align}
None-zero elements of $\tilde{\textbf{H}}_M$ are listed below (the Hamiltonian is a Hermitian matrix)
\begin{align}
(\tilde{\textbf{H}}_M)_{14}&=-\frac{1}{2E}\,\big(a+b\,c^2_{24}c^2_{34}\big)\,c_{14}s_{14}, \notag \\
(\tilde{\textbf{H}}_M)_{24}&=-\frac{b}{2E}\,c_{14}c_{24}s_{24}c^2_{34}\,e^{i\delta_{24}}, \notag \\
(\tilde{\textbf{H}}_M)_{34}&=-\frac{b}{2E}\,c_{14}c_{24}c_{34}s_{34}\,e^{i\delta_{34}}. \notag \\
(\tilde{\textbf{H}}_M)_{44}&=0.
\end{align}
Since $s_{i4}\sim\mathcal{O}(\sqrt{\epsilon})$, it is easy to see that $\tilde{\textbf{H}}_M\sim\mathcal{O}(\sqrt{\epsilon})$. Although $\tilde{\textbf{H}}_M$ is not as small as $\mathcal{O}(\epsilon)$, it will be a part of the perturbative Hamiltonian. However, this does not mean that the first order corrections must be as large as $\mathcal{O}(\sqrt{\epsilon})$. The mass of the heavy sterile neutrino will be an alternative parameter which controls scales of the correction terms. More specifically, in a perturbative expression, all non-zero elements of $\tilde{\textbf{H}}_M$ are divided by $M^2$. For large $M^2$ the quotient gives a small term in the perturbation expansion. Another condition that is necessary for $\tilde{\textbf{H}}_M$ being a perturbative Hamiltonian is that it consists of terms proportional to $a$ and $b$,  which means that it vanishes in vacuum. This is crucial because we require the perturbative expressions to be exact in vacuum. 
 
\begin{figure}[t]
\centering
\includegraphics[width=0.4\textwidth]{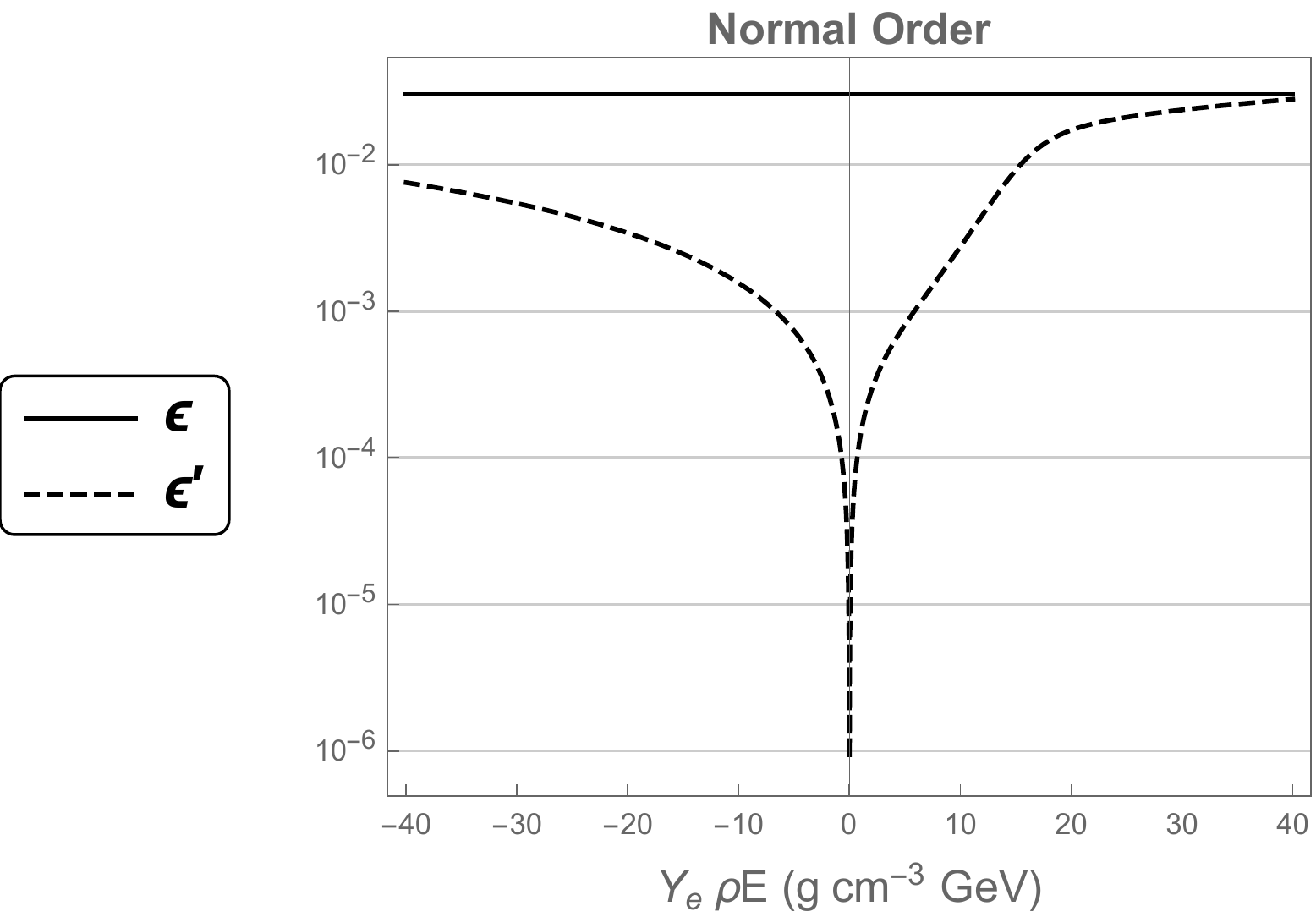}
\caption{The perturbing parameter $\epsilon^\prime$ as function of  $Y_e\,\rho E$ with $b= a/2$.
In the region where $a$ is comparable to $\Delta m^2_{ee}$,  $\epsilon^\prime \leq \epsilon $. The parameters used are in Table\,\ref{table:parameters}. }
\label{fig:Epsilon}
\end{figure}

\subsection{$\textbf{U}_{13}$ rotation}
Now the dominating off-diagonal term (except the ones in $\tilde{\textbf{H}}_M$) comes from the (1-3) sector of $\tilde{H}_0$. Because of the complex phase $\delta_{34}$, the rotation will not be real. Let us assume that the rotation is $\textbf{U}_{13}(\tilde{\theta}_{13},\alpha_{13})$, where $\tilde{\theta}_{13}$ is the real rotation angle and $\alpha_{13}$\footnote{Here we are not using the usual phase symbol $\delta$ since $\alpha_{13}$ is not an effective physical phase in matter. In Appendix \ref{appendix:phase} it can be eliminated by implementing a pure phase transformation of the neutrino basis.} is the complex phase. After this rotation the neutrino basis becomes
\begin{align}
\ket{\hat{\nu}}&\equiv \textbf{U}^\dagger_{13}(\tilde{\theta}_{13},\alpha_{13})\ket{\tilde{\nu}} \notag \\
&=\textbf{U}^\dagger_{13}(\tilde{\theta}_{13},\alpha_{13})\textbf{U}^\dagger_\text{sterile}\textbf{U}^\dagger_{23}(\theta_{23},\delta_{23})\ket{\nu}_\text{f},
\end{align}
where $\textbf{U}^\dagger_\text{sterile}=\textbf{U}^\dagger_{14}(\theta_{14})\textbf{U}^\dagger_{24}(\theta_{24},\delta_{24})\textbf{U}^\dagger_{34}(\theta_{34},\delta_{34})$. The Hamiltonian becomes
\begin{align}
\hat{\textbf{H}}&\equiv \textbf{U}^\dagger_{13}(\tilde{\theta}_{13},\,\alpha_{13})\,\tilde{\textbf{H}}\,\textbf{U}_{13}(\tilde{\theta}_{13},\alpha_{13}) .
\end{align}
Since the 4th index is not engaged in the rotation, we can just focus on the first three indices and define a $3\times 3$ submatrix $U_{13}$ to be the active sectors of $\textbf{U}_{13}$, i.e.
\begin{equation}
\textbf{U}_{13}=\left(\begin{array}{cc}
U_{13}& \\
& 1
\end{array}\right).
\end{equation}
After the rotation, the sub-Hamiltonian in the active sector $\tilde{H}$ becomes
\begin{equation}
\hat{H}\equiv U^\dagger_{13}(\tilde{\theta}_{13},\alpha_{13})\,\tilde{H}\,U_{13}(\tilde{\theta}_{13},\alpha_{13}).
\end{equation}
We require the (1-3) sector of $\tilde{H}$ to be diagonalized by $U_{13}(\tilde{\theta}_{13},\alpha_{13})$. Since the (1-3) sector of $\tilde{H}_1$ vanishes,  it is equivalent to diagonalizing this sector of $\tilde{H}_0$, i.e. 
\begin{align}
\hat{H}_0&\equiv U^\dagger_{13}(\tilde{\theta}_{13},\alpha_{13})\,\tilde{H}_0\,U_{13}(\tilde{\theta}_{13},\alpha_{13}) \notag \\
&=\frac{1}{2E}\left(\begin{array}{ccc}
\lambda_- & & \\
 &\lambda_0 & \\
 & & \lambda_+
\end{array}\right),
\end{align}
with $\lambda_\pm$ and $\lambda_0$ to be determined. Simultaneously $\tilde{H}_1$ becomes
\begin{equation}
\hat{H}_1\equiv U^\dagger_{13}(\tilde{\theta}_{13},\alpha_{13})\,\tilde{H}_1\,U_{13}(\tilde{\theta}_{13},\alpha_{13}).
\end{equation}
It can be shown that
\begin{widetext}
\begin{align}
 \lambda_\mp&=\frac{1}{2}\bigg[\,(\lambda_a+\lambda_c)\mp\text{sign}(\Delta m^2_{ee})\sqrt{(\lambda_c-\lambda_a)^2+4\left|s_{13}c_{13}\,\Delta m^2_{ee}+\epsilon\,b\,k_{13}\,c_{24}c_{34}\,e^{-i\delta_{34}}\right|^2}\,\bigg], \notag \\
 \lambda_0&=\lambda_b=\epsilon\,c^2_{12}\,\Delta m^2_{ee}+\epsilon\,b\,k_{22}\,c^2_{34}.
\end{align}
\end{widetext}
The real rotation angle and the complex phase can be determined by
\begin{align}
 \cos{2\tilde{\theta}_{13}}&=\frac{\lambda_c-\lambda_a}{\lambda_+-\lambda_-}, \notag \\
 \alpha_{13}&=\text{Arg}\big[s_{13}c_{13}\,\Delta m^2_{ee}+\epsilon\,b\,k_{13}\,c_{24}c_{34}\,e^{-i\delta_{34}}\big].
\end{align}
The elements of $\hat{H}_1$ are
\begin{align}
(\hat{H}_1)_{12}&=\frac{\epsilon}{2E}\notag \\
\times\bigg\{&c_{12}s_{12}\,(\,c_{13}\tilde{c}_{13}+s_{13}\tilde{s}_{13}\,e^{-i\,\alpha_{13}}\,)\,\Delta m^2_{ee} \notag \\
 +b&\,\Big[\,k_{12}\,c_{24}c^2_{34}\tilde{c}_{13}-k_{23}\,c_{34}\tilde{s}_{13}\,e^{i(\delta_{34}+\alpha_{13})}\,\Big]\,e^{-i\delta_{24}}\,\bigg\}, \notag \\
 (\hat{H}_1)_{23}&=\frac{\epsilon}{2E}\notag \\
 \times\bigg\{&c_{12}s_{12}\,(\,-s_{13}\tilde{c}_{13}+c_{13}\tilde{s}_{13}\,e^{i\,\alpha_{13}}\,)\,\Delta m^2_{ee} \notag \\
 +b&\,\Big(\,k_{12}\,c_{24}c^2_{34}\tilde{s}_{13}\,e^{i\,\alpha_{13}}+k_{23}\,c_{34}\tilde{c}_{13}\,e^{-i\delta_{34}}\,\Big)\,e^{i\delta_{24}}\,\bigg\}, \notag \\
 (\hat{H}_1)_{13}&=0. \label{eq:hhat}
\end{align}

The Hamiltonian in the sterile sector becomes 
\begin{equation}
\hat{\textbf{H}}_M\equiv \textbf{U}^\dagger_{13}(\tilde{\theta}_{13},\alpha_{13})\,\tilde{\textbf{H}}_M\,\textbf{U}_{13}(\tilde{\theta}_{13},\alpha_{13}).
\end{equation}

At the end of this subsection we define a real parameter $\epsilon^\prime$ and a phase $\alpha_\epsilon$ 
\begin{align}
\epsilon^\prime&\equiv\bigg|\frac{2E}{\Delta m_{ee}^2}(\hat{H}_1)_{23}\bigg|, \notag \\
\alpha_\epsilon&\equiv\text{Arg}\big[\frac{2E}{\Delta m_{ee}^2}(\hat{H}_1)_{23}\big].
\end{align}
Obviously $\epsilon^\prime\sim\epsilon$ and $(\hat{H}_1)_{23}=e^{i\alpha_\epsilon}\epsilon^\prime\Delta m^2_{ee}/2E$. It is not hard to see that in the Standard Model $\epsilon^\prime=|\epsilon\,\sin{(\tilde{\theta}_{13}-\theta_{13})}\,s_{12}c_{12}|$, which reconciles with the one defined in \cite{Denton:2016wmg}. The two new defined parameters will frequently emerge in the following sections. Since in vacuum, $a,\,b=0$, $\tilde{\theta}_{13}=\theta_{13}$ and $\alpha_{13}=0$, $\epsilon^\prime$ must be zero then, as shown in Fig.\,\ref{fig:Epsilon}. This guarantees that the perturbative expressions will be exact in vacuum.

\begin{figure*}[t]
\centering
\includegraphics[width=2.265in]{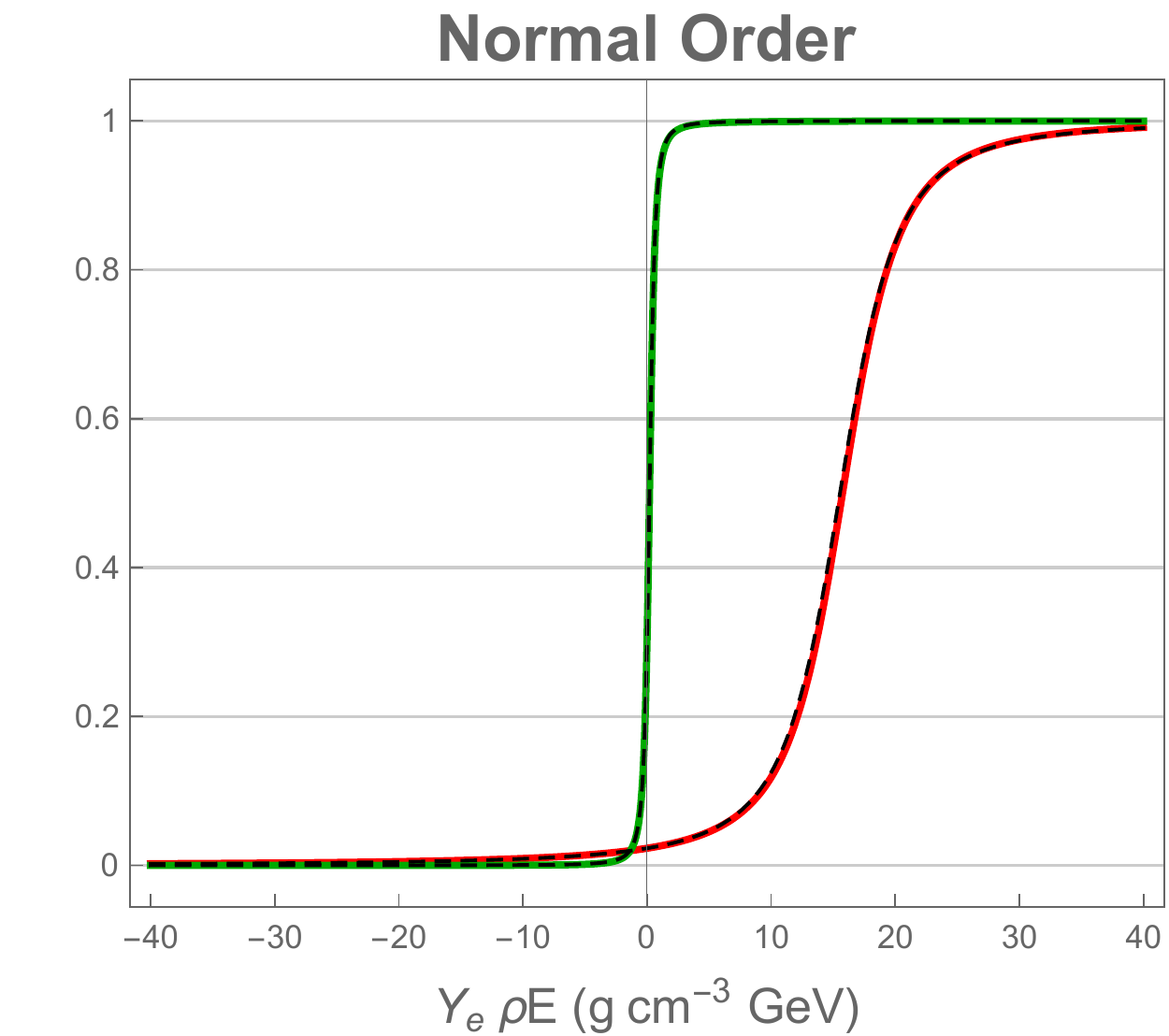}
\includegraphics[width=3.235in]{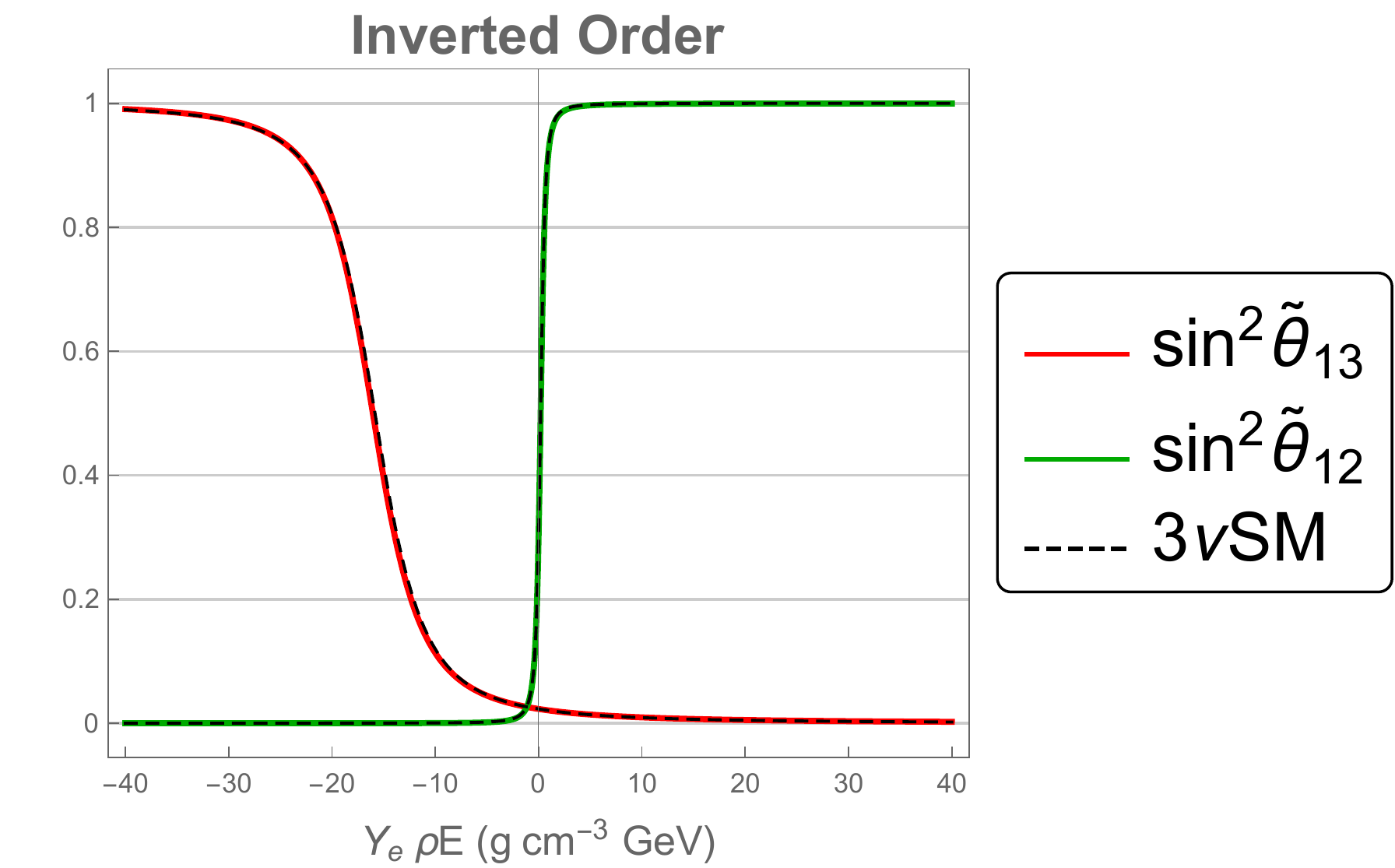}
\caption{Values of $\sin^2\tilde{\theta}_{13}$ and $\sin^2\tilde{\theta}_{12}$. The solid lines are values in 3+1 scheme; as a comparison the dashed lines are the values in $3\nu SM$. The differences are small but non-negligible.  The parameters used are in Table\,\ref{table:parameters}. }
\label{fig:angles}
\end{figure*}

\subsection{$\textbf{U}_{12}$ rotation}
As pointed out in \cite{Denton:2016wmg}, to resolve the $\lambda_1$ and $\lambda_0$ crossing at the solar resonance, one more rotation that diagonalizes the (1-2) sector is necessary. Again, since $(\hat{H}_1)_{12}$ is complex, the rotation cannot be real in general. We assume that the rotation in (1-2) sector is $\textbf{U}_{12}(\tilde{\theta}_{12},\alpha_{12})$, and after this rotation, the neutrino basis becomes

\begin{align}
\ket{\check{\nu}}\equiv&\textbf{U}^\dagger_{12}(\tilde{\theta}_{12},\alpha_{12})\ket{\hat{\nu}} \notag \\
=&\textbf{U}^\dagger_{12}(\tilde{\theta}_{12},\alpha_{12})\textbf{U}^\dagger_{13}(\tilde{\theta}_{13},\alpha_{13})\textbf{U}^\dagger_\text{sterile}\textbf{U}^\dagger_{23}(\theta_{23},\delta_{23})\ket{\nu}_\text{f},
\end{align}
where $\textbf{U}_\text{sterile}=\textbf{U}^\dagger_{14}(\theta_{14})\textbf{U}^\dagger_{24}(\theta_{24},\delta_{24})\textbf{U}^\dagger_{34}(\theta_{34},\delta_{34})$
The Hamiltonian becomes
\begin{equation}
\check{\textbf{H}}\equiv \textbf{U}^\dagger_{12}(\tilde{\theta}_{12},\alpha_{12})\,\hat{\textbf{H}}\,\textbf{U}_{12}(\tilde{\theta}_{12},\alpha_{12}).
\end{equation}
Similar to the case of the (1-3) rotation, we can again define a $3\times 3$ submatrix $U_{12}$ by
\begin{equation}
\textbf{U}_{12}=\left(\begin{array}{cc}
U_{12}& \\
& 1
\end{array}\right).
\end{equation}
Now we require the $U_{12}(\tilde{\theta}_{12},\alpha_{12})$ to diagonalize the (1-2) sector of $\hat{H}$. After the rotation the sub-Hamiltonian is
\begin{align}
\check{H}&\equiv U^\dagger_{12}(\tilde{\theta}_{12},\alpha_{12})\,\hat{H}\,U_{12}(\tilde{\theta}_{12},\alpha_{12}) \notag \\
&=\check{H}_0+\check{H}_1,
\end{align}
where $\check{H}_0$ and $\check{H}_1$ are in zeroth and first order respectively, i.e.

\begin{align}
\check{H}_0&=\frac{1}{2E}\left(\begin{array}{ccc}
\lambda_1& & \\
&\lambda_2 & \\
& &\lambda_3 \\
\end{array}\right), \notag \\
\check{H}_1&=\frac{\epsilon^\prime\Delta m^2_{ee}}{2E} \notag \\
\times&\left(\begin{array}{ccc}
& &\,-\tilde{s}_{12}\,e^{i(\alpha_{12}+\alpha_\epsilon)} \\
& &\,\tilde{c}_{12}\,e^{i\alpha_\epsilon} \\
-\tilde{s}_{12}\,e^{-i(\alpha_{12}+\alpha_\epsilon)} & \quad\tilde{c}_{12}\,e^{-i\alpha_\epsilon} &
\end{array}\right).
\end{align}

The diagonal elements of $\check{H}_0$ are
\begin{align}
\lambda_{1,2}&=\frac{1}{2}\bigg[\,(\lambda_-+\lambda_0)\mp\sqrt{(\lambda_--\lambda_0)^2+4|(\hat{H}_1)_{12}|^2}\,\bigg], \notag \\
\lambda_3&=\lambda_+. \label{eq:lambda123}
\end{align}

The real rotation angle and the complex phase can be determined by
\begin{align}
\cos{2\tilde{\theta}_{12}}&=\frac{\lambda_0-\lambda_-}{\lambda_2-\lambda_1}, \notag \\
\alpha_{12}&=\text{Arg}\big[(\hat{H}_1)_{12}\big].
\end{align}

Values of $\sin^2\tilde{\theta}_{13}$ and $\sin^2\tilde{\theta}_{12}$ are plotted in Fig.\,\ref{fig:angles}. After this (1-2) rotation, crossings of the first two diagonal elements $\lambda_{1,2}$ have been resolved, as shown in the top panels of Fig.\,\ref{fig:lambda}.  They will be the zeroth order eigenvalues in the following perturbation expansions in the next section. The difference between 3+1 and $3\nu$SM is small in both panels of  Fig.\,\ref{fig:angles} and the bottom panels of Fig.\,\ref{fig:lambda} but not insignificant.

The Hamiltonian in the sterile sector now is
\begin{equation}
\check{\textbf{H}}_M\equiv \textbf{U}^\dagger_{12}(\tilde{\theta}_{12},\alpha_{12})\,\hat{\textbf{H}}_M\,\textbf{U}_{12}(\tilde{\theta}_{12},\alpha_{12}).
\end{equation}
From $\tilde{\textbf{H}}_M$ to $\check{\textbf{H}}_M$, we implemented two rotations in (1-3) and (1-2) sectors. Because the active and sterile sectors were not mixed by the two rotations, the elements are still combinations of the terms proportional to $s_{i4}\sim\mathcal{O}(\sqrt{\epsilon})$. Elements of $\check{\textbf{H}}_M$ can be found in Appendix \ref{appendix:checkHM}.

\subsection{Crossings of $M^2$}
\begin{figure*}[t]
\centering
\includegraphics[width=2.455in]{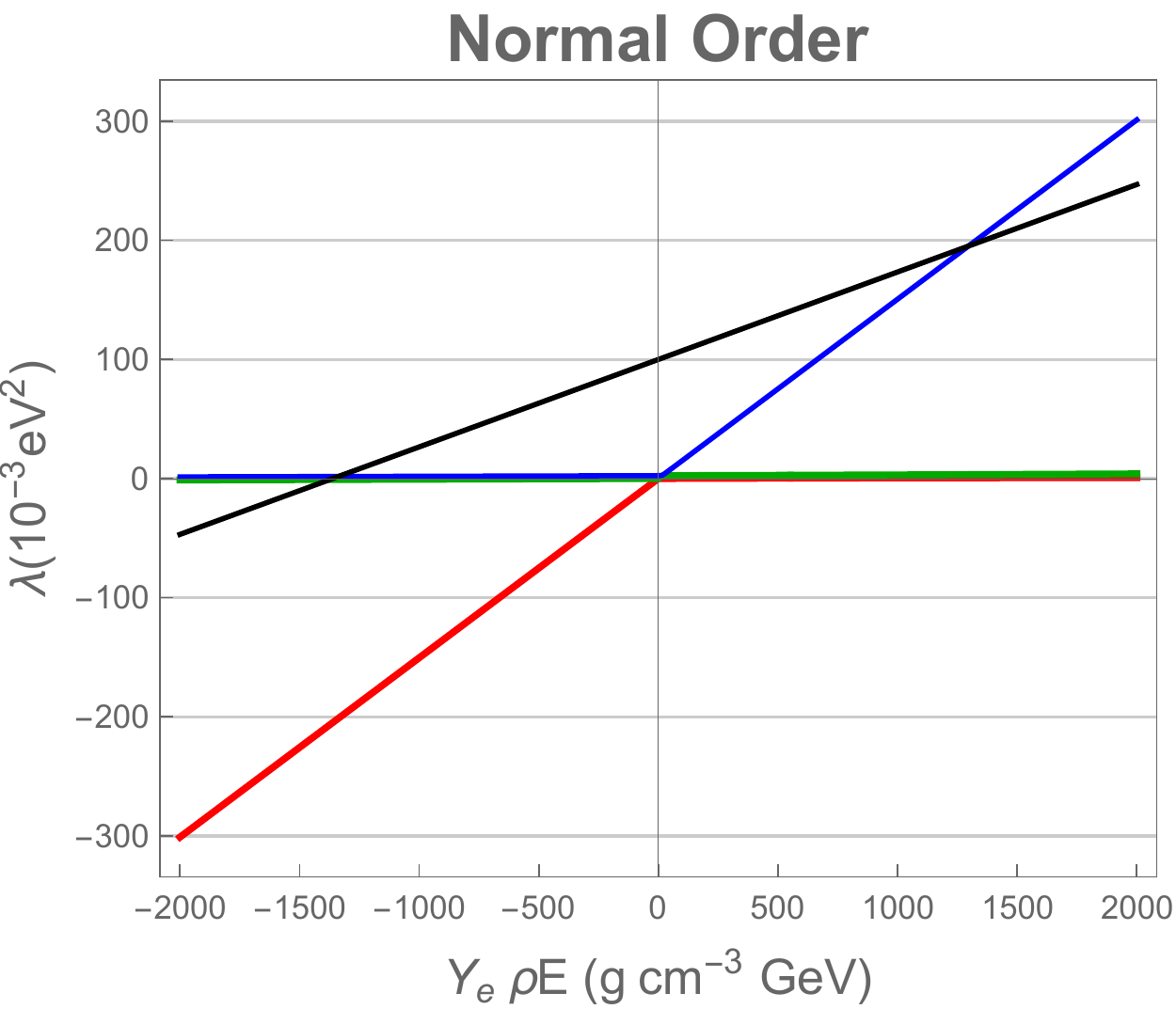}
\includegraphics[width=3.045in]{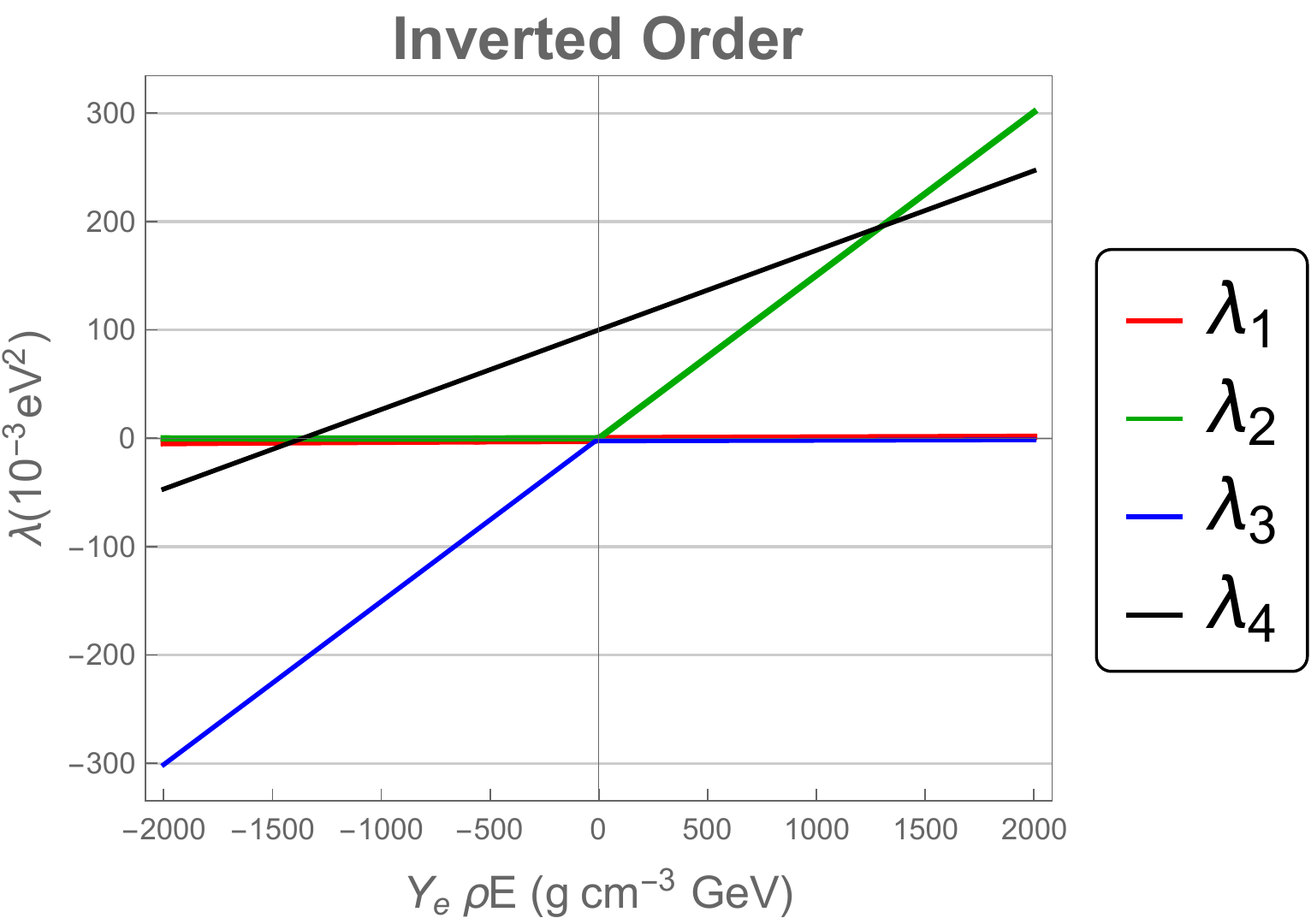}
\includegraphics[width=2.39in]{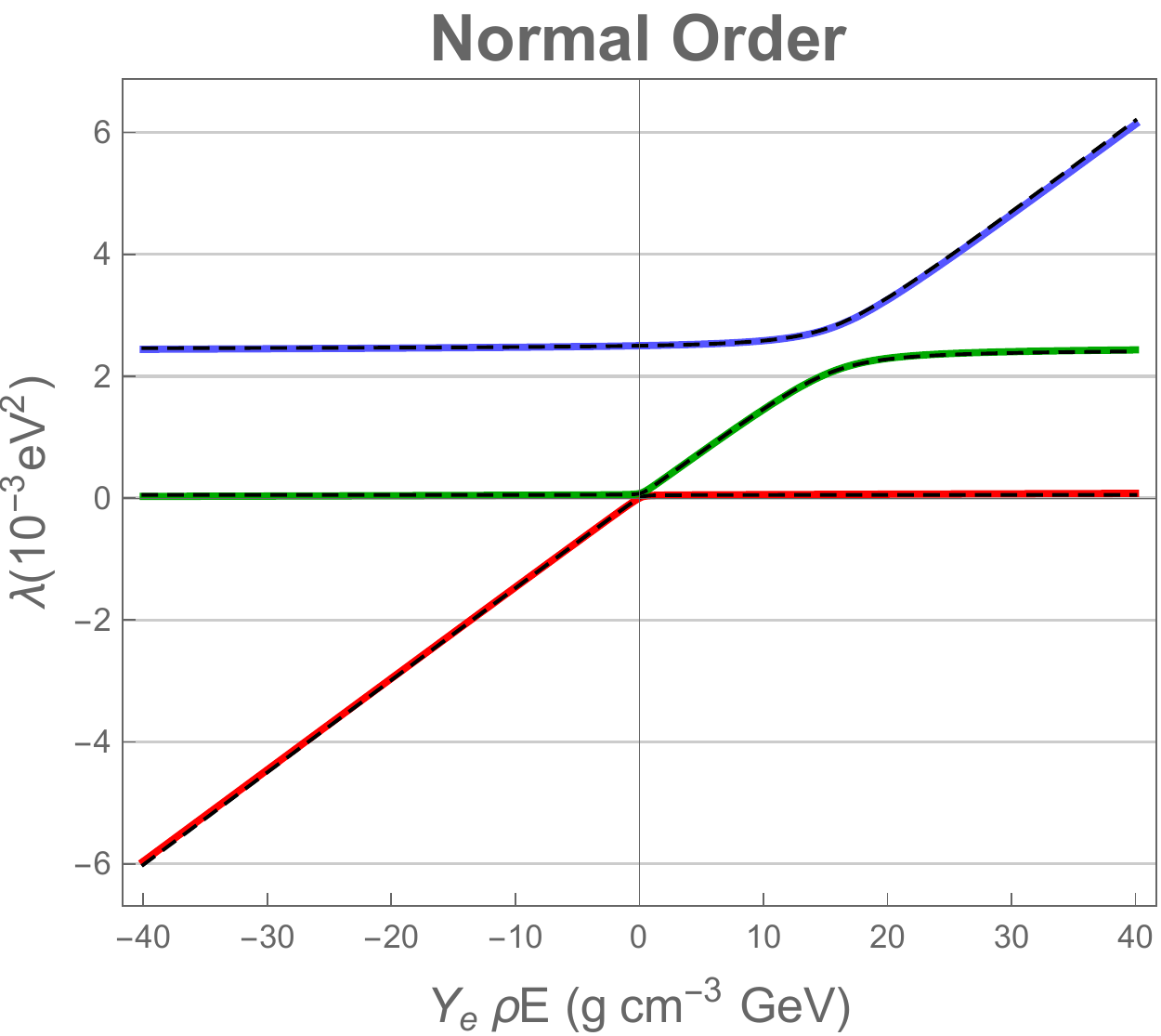}
\includegraphics[width=3.21in]{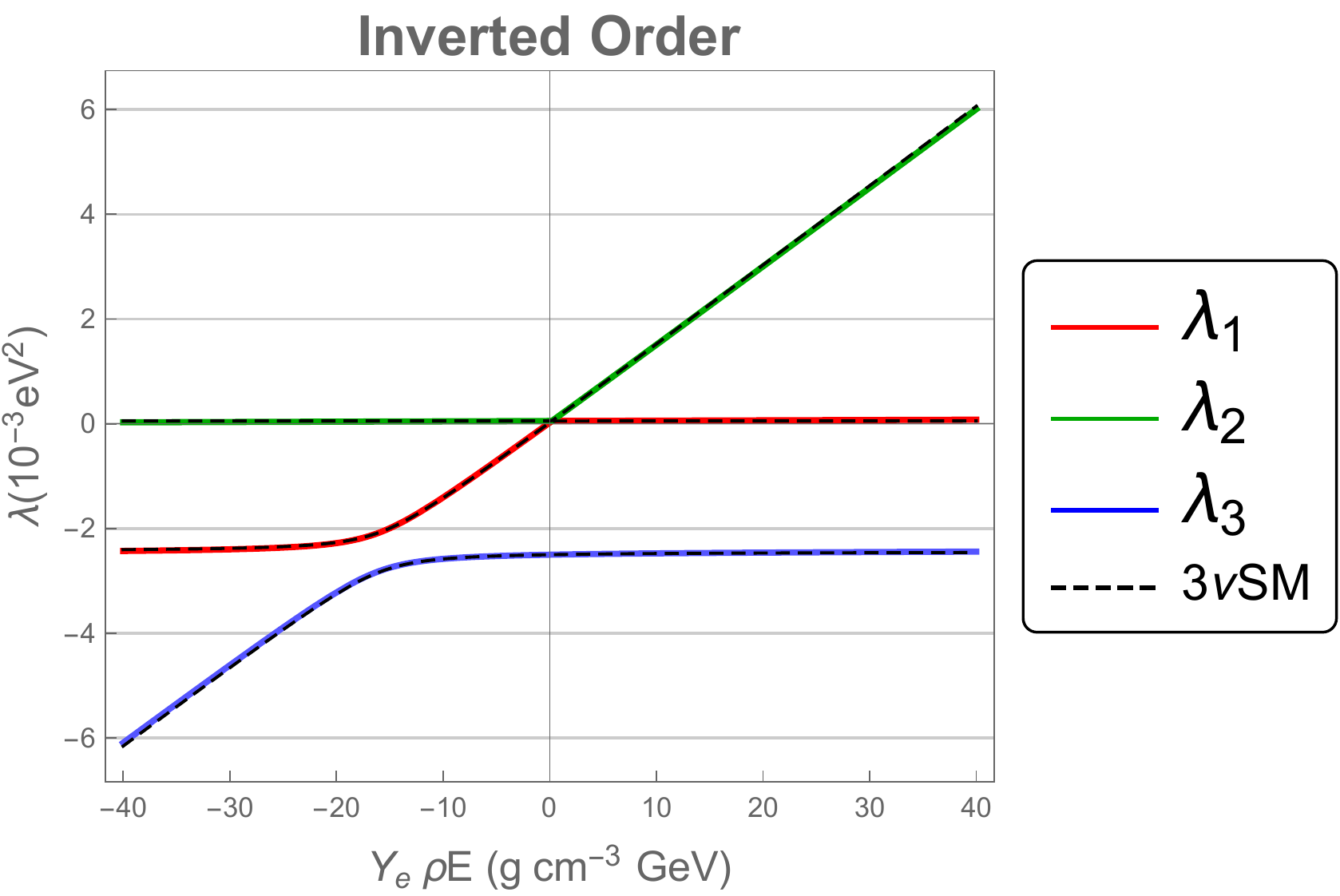}
\caption{The top two panels give the crossing of the fourth eigenvalue (black), using $\Delta m^2_{41}=0.1\,\text{eV}^2$, with the active eigenvalues (red, green and blue). The active eigenvalues, $\lambda_{1,2,3}$ can cross $\lambda_4=M^2(b)$ only if the neutrino energy is very large (${\cal O}(1)$ TeV for earth densities). The bottom two panels are zoomed in to the region of primary interest; they show the zeroth order active eigenvalues in normal and inverted order; also for comparison, the dashed lines are the values in $3\nu SM$.  Again the differences are small but non-negligible. The parameters used are in Table\,\ref{table:parameters}.}
\label{fig:lambda}
\end{figure*}

In principle, there are still some possible crossings of the diagonal elements, namely the crossings to the fourth diagonal element. Since both the (1-3) and the (1-2) rotations are in the active space (first three rows and columns), the fourth element is still 
\begin{equation}
M^2(b )    \equiv \Delta m^2_{41}+\,b\,c^2_{14}c^2_{24}c^2_{34},
\end{equation}
since $\Delta m^2_{41}$ is much larger than the active eigenvalues in vacuum. Thus, the crossings to $M^2$ can only happen with very high neutrino energy, as shown in the top panels  of Fig.\,\ref{fig:lambda}. From the figure we can see that if $Y_e\rho= 1.4\,\text{g}\,\cdot\text{cm}^{-3}$, for the earth's crust, the neutrino energy must be ${\cal O}(1)\,\text{TeV}$. Considering the energy scales of the current and future accelerator based oscillation experiments, we are
therefore not considering the energy region of these additional crossings,  so they will not effect our result. For much higher energy experiments these additional level crossings would have to be dealt with using matter additional rotations.

\subsection{Summary of the rotations} 
Now $\check{H}_0$'s diagonal elements, $\lambda_{1,2,3}$, do not cross (crossings to $M^2$ will not happen in the energy region of interest). All the off-diagonal elements in the active sectors are of scale $\epsilon^\prime$. We will distribute all the diagonal elements to the zeroth order Hamiltonian and all the off-diagonal elements to the perturbative Hamiltonian, i.e.
\begin{equation}
    \check{\textbf{H}}_0=\left(\begin{array}{cc}
        \check{H}_0 &  \\
         & \frac{M^2}{2E}
  \end{array}\right),\quad\check{\textbf{H}}_1=\left(\begin{array}{cc}
        \check{H}_1 &  \\
         & 0
    \end{array}\right)+\check{\textbf{H}}_M.
\end{equation}
The zeroth order effective PMNS matrix in matter is

\begin{align}
&\textbf{U}^m_\text{PMNS} \notag \\ =&\textbf{U}_{23}(\theta_{23},\delta_{23})\textbf{U}_{34}(\theta_{34},\delta_{34})\textbf{U}_{24}(\theta_{24},\delta_{24})\textbf{U}_{14}(\theta_{14})\notag \\
&\quad\times\textbf{U}_{13}(\tilde{\theta}_{13},\alpha_{13})\textbf{U}_{12}(\tilde{\theta}_{12},\alpha_{12}). \label{eq:umpmns}
\end{align}

\begin{figure*}[t]
\centering
\includegraphics[width=5in]{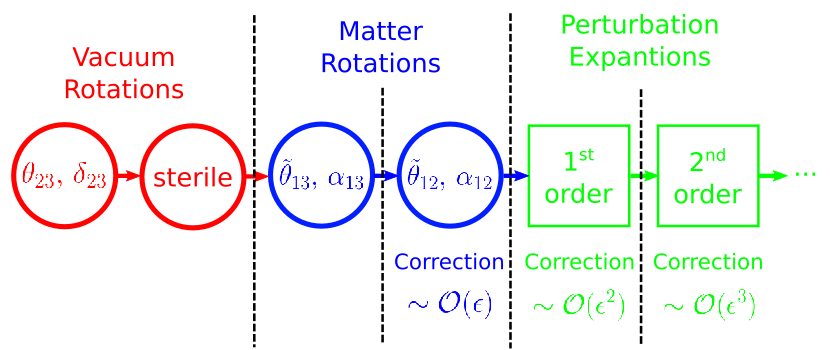}
\caption{Summary of the rotations and the following perturbative expansions. We first implemented vacuum rotations in the (2-3) and sterile sectors, the red circle with text \emph{sterile} inside indicates the rotations in sterile rotations, i.e. the rotations represented by $U_\text{sterile}=U_{34}\,U_{24}\,U_{14}$, see Eq.\,\ref{eq:Usterile}; then two matter rotations in (1-3) and (1-2) sectors were performed. After the series of rotations, the zeroth order approximations of the eigenvalues and eigenvectors achieve $\mathcal{O}(\epsilon)$ accuracy. Perturbative expansions will be used to further improve the precision.}
\label{fig:flowchart}
\end{figure*}

Since all possible degeneracies have been removed in the energy scale which we are interested in, we are free to implement a perturbation expansion to achieve even better accuracy. The process of reducing errors by performing rotations and perturbative expansions is summarized in Fig.\,\ref{fig:flowchart}.

For the scenario with more than one sterile neutrino, although it is more complicated, the rotation method developed here is still applicable. Using the same convention of Eq.\,\ref{eq:convention} to define the PMNS matrix and implement the rotations in the sequence of $U_{23}\rightarrow U_\text{sterile}\rightarrow U_{13}\rightarrow U_{12}$ as above.

\section{Perturbative expressions} \label{sec:perturbation}

Since all the crossings of the zeroth order eigenvalues have been resolved (except for the crossings with $M^2$, which are not in the energy region of interest) by the rotations and all the off-diagonal elements are small, we can now calculate the higher order corrections to the eigenvalues and eigenvectors by perturbation methods . 

We define $\textbf{V}$ to be the exact PMNS matrix in matter. It can be related to the zeroth order $\textbf{U}^m_\text{PMNS}$ by 
\begin{equation}
\textbf{V}=\textbf{U}^m_\text{PMNS}(\mathbb{1}+\textbf{W}_1+\textbf{W}_2+\cdots), \label{eq:V}
\end{equation}
where $\textbf{W}_n$ is $n$th order correction. The exact eigenvalues are
\begin{equation}
    \lambda^{(ex)}_i=\lambda_i+\lambda_i^{(1)}+\lambda_i^{(2)}+\cdots,\quad i=1,2,3,4\,\,\,,
\end{equation}
where $\lambda_{1,2,3}$ are defined in Eq.\,\ref{eq:lambda123} and $\lambda_4=M^2$, $\lambda^{(n)}_i$ is the $n$th order correction.

First order corrections to the eigenvalues are
\begin{equation}
\lambda_i^{(1)}=2E(\check{\textbf{H}}_1)_{ii}=0. \label{eq:lambdai1}
\end{equation}
First order corrections to the eigenstates are determined by $\textbf{W}_i$ defined in Eq.\,\ref{eq:V}, which are
\begin{equation}
(\textbf{W}_1)_{ij}=\begin{cases}
0 & i=j \\
-\frac{2E(\check{\textbf{H}}_1)_{ij}}{\lambda_i-\lambda_j} & i\neq j
\end{cases}. \label{eq:w1}
\end{equation}
The detailed first and second order formulas of the perturbation expansions can be found in Appendix \ref{appendix:perturbation}. In general, with crossings of the zeroth order eigenvalues ruled out, perturbative expansions can go to arbitrary precision. However, numerical tests will suggest that it is sufficient to terminate the approach at second order.  

\begin{figure*}[t]
\centering
\includegraphics[width=6in]{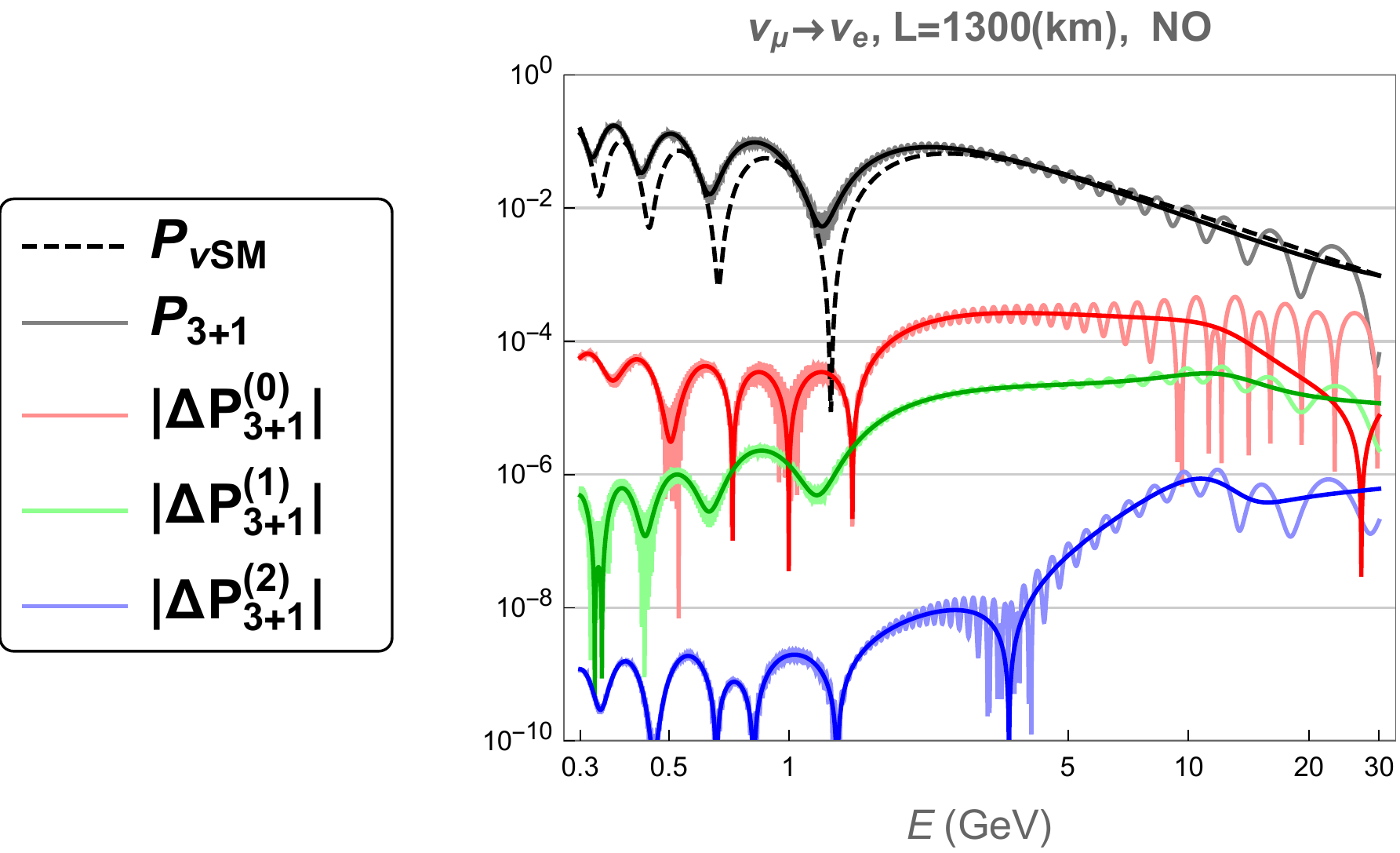}
\includegraphics[width=6in]{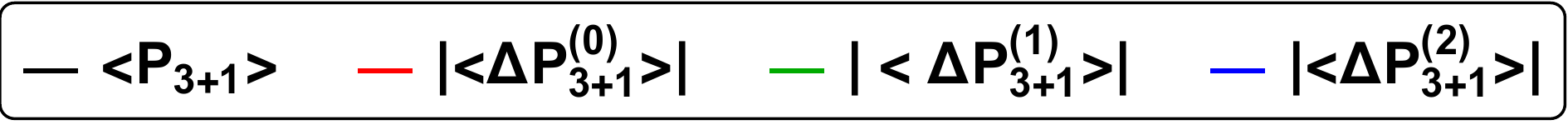}
\caption{In the 3+1 scheme, errors of the zeroth, first and second order approximations are presented by red, green and blue curves, respectively. The light colors (which look like bold shadows in low energy region) are representing true corrections; the darker ones are showing the expectation values. The exact probability (expectation value) in the 3+1 scheme, which is plotted by the gray solid (black solid) curve, can be calculated by \cite{Li:2018ezt}. As a contrast, the dashed black line is showing the probabilities in the Standard Model, with $Y_e\rho=1.4\,\text{g}\cdot\text{cm}^{-3}$.}
\label{fig:P_DeltaP}
\end{figure*}

\subsection{Numerical precision test}\label{subsec:numericaltesst}
We now test the accuracy of our perturbative expressions. We choose the $\nu_{\mu}\rightarrow\nu_e$ channel and 1300km baseline of DUNE to do the numerical test. The density of the earth crust is chosen to be $Y_e\rho=1.4\,\text{g}\cdot\text{cm}^{-3}$, $b=a/2$ and all the mixing parameters are listed in Table\,\ref{table:parameters}. The exact oscillation probabilities can be figured out by \cite{Li:2018ezt} or given by a computer algebra system\footnote{Only considering the 3+1 scheme, an analytical solution is still possible since one just need solve a quartic equation; but it is not the case for schemes with more sterile neutrinos}. The results are shown in Fig.\,\ref{fig:P_DeltaP}. The error in the zeroth order expression is expected to be no more than $\epsilon\sim 10^{-2}$, which is confirmed by the red curve in the plot; the green curve depicts the error of the first order perturbative expansion, which is under $\epsilon^2\sim10^{-4}$; to second order, the error further declines to $\epsilon^3\sim10^{-6}$, which also coincides with the prediction. In Fig.\,\ref{fig:P_DeltaP} the expectation values are obtained by averaging over the fast oscillation terms, i.e. the terms with angular velocities proportional to $(\lambda_4-\lambda_i)$. More specifically, 
\begin{equation}
\big<\sin/\cos\frac{(\lambda_4-\lambda_i)L}{2E}\big>=0,\,\, \big<\sin^2\frac{(\lambda_4-\lambda_i)L}{4E}\big>=\frac{1}{2}.
\end{equation}
Based on the numerical results, we confirm that at least the second order perturabtive expansion is significantly more accurate than any experimental results \cite{Acciarri:2015uup,Patterson:2012zs,Abe:2011ks,Abe:2015zbg,Kelly:2018kmb}.

\begin{figure*}[t]
\centering
\includegraphics[width=3in]{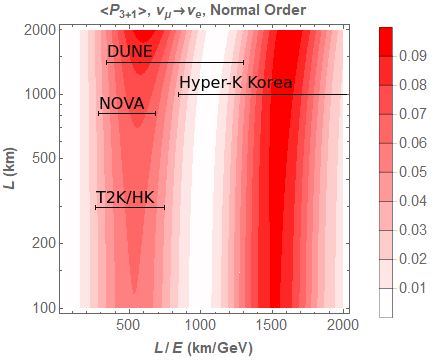}
\includegraphics[width=3in]{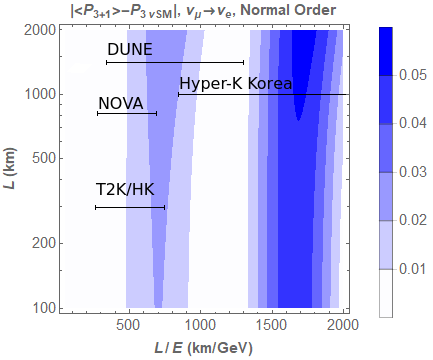}
\caption{For $\nu_\mu\rightarrow\nu_e$ channel, the left plot is showing the probabilities predicted by the 3+1 scheme; differences of the probabilities (expectation values, with fast oscillations averaged out) predicted by the standard three-flavor scheme and the 3+1 scheme are presented in the right plot. $P_{3+1}$ in both figures are computed by the 0th order rotation method developed in this paper. Parameters used are given in Table \ref{table:parameters}. Neutrino flux energies used are $0.4-1.2\,\text{GeV}$ for T2K/HyperK (295 km), $1.2-3.0\,\text{GeV}$ for NOVA (810 km), $0.4-1.5\,\text{GeV}$ for T2HKK (1100 km) and $1.0-4.0\,\text{GeV}$ for DUNE (1300 km), see \cite{Abe:2016aoo,Acero:2019qcr,Papadimitriou:2018akk,Abe:2016ero}.}
\label{fig:Psm_Psterile_mue}
\end{figure*}

\begin{figure*}[t]
\centering
\includegraphics[width=2.9in]{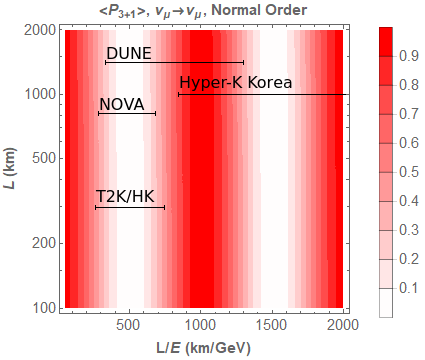}
\includegraphics[width=2.9in]{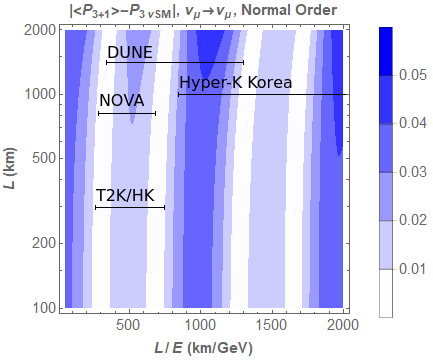}
\caption{For $\nu_\mu\rightarrow\nu_\mu$ channel, the left plot is showing the probabilities (expectation values, with fast oscillations averaged out) predicted by the 3+1 scheme; differences of the probabilities predicted by the standard three-flavor scheme and the 3+1 scheme are presented in the right plot. $P_{3+1}$ in both figures are computed by the 0th order rotation method developed in this paper. Parameters used are given in Table \ref{table:parameters}. See Fig.\,\ref{fig:Psm_Psterile_mue} for neutrino flux energies of the listed facilities.}
\label{fig:Psm_Psterile_mumu}
\end{figure*}

\begin{figure*}[t]
\centering
\includegraphics[width=3in]{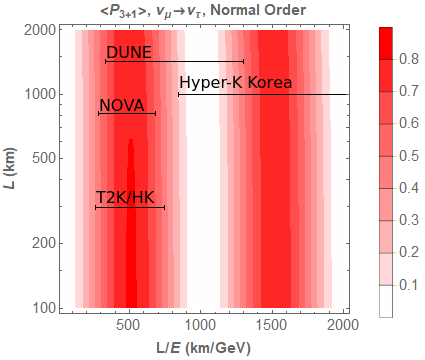}
\includegraphics[width=3in]{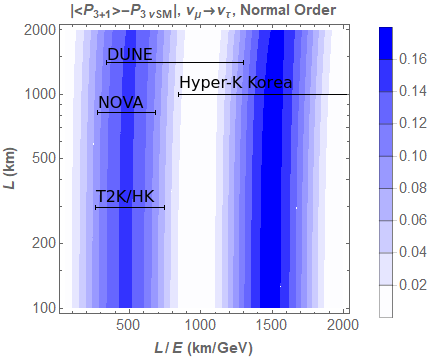}
\caption{For $\nu_\mu\rightarrow\nu_\tau$ channel, the left plot is showing the probabilities (expectation values, with fast oscillations averaged out) predicted by the 3+1 scheme; differences of the probabilities predicted by the standard three-flavor scheme and the 3+1 scheme are presented in the right plot. $P_{3+1}$ in both figures are computed by the 0th order rotation method developed in this paper. Parameters used are given in Table \ref{table:parameters}. See Fig.\,\ref{fig:Psm_Psterile_mue} for neutrino flux energies of the listed facilities.}
\label{fig:Psm_Psterile_mutau}
\end{figure*}

\begin{table*}
\begin{center}
\begin{tabular}{|p{3cm}|p{0.7cm}|p{0.7cm}|p{0.7cm}|p{1cm}|p{0.7cm}|p{0.7cm}|p{1cm}|p{0.7cm}|p{1cm}|}
\hline
\centering{$U_\text{PMNS}\equiv$}&\centering{$s^2_{12}$}&\centering{$s^2_{13}$}&\centering{$s^2_{23}$}&\centering{$\delta_{23}/\pi$}&\centering{$s^2_{14}$}&\centering{$s^2_{24}$}&\centering{$\delta_{24}/\pi$}&\centering{$s^2_{34}$}&$\delta_{34}/\pi$\\ 
\hline
$U_\text{sterile}\,U_{23}\,U_{13}\,U_{12}$&\multirow{2}{*}{0.3}&\multirow{2}{*}{0.02}&0.44&-0.40&\multirow{2}{*}{0.02}&0.01&0.10&0.1&\quad0\\
\cline{1-1}\cline{4-5}\cline{7-10}
$U_{23}\,U_\text{sterile}\,U_{13}\,U_{12}$& & &0.49&-0.39& &0.02&0.50&0.09&0.08\\
\hline
\end{tabular}
\caption{Mixing parameters and vacuum eigenvalues used for the numerical calculations \cite{Fernandez-Martinez:2016lgt,Kopp:2013vaa,deGouvea:2015euy,TheIceCube:2016oqi}. In different conventions to define the PMNS matrix (orders of $U_{23}$ and $U_\text{sterile}$, where $U_\text{sterile}=U_{34}\,U_{24}\,U_{14}$, see Eq.\,\ref{eq:Usterile}), some of the parameters are different, formulas to relate the parameters in both conventions are in Appendix \ref{appendix:convention}. In both conventions the energy eigenvalues in vacuum are $\Delta m^2_{21}=7.5\times10^{-5}\,\text{eV}^2$, $\Delta m^2_{31}=2.5\times10^{-3}\,\text{eV}^2$ and $\Delta m^2_{41}=0.1\,\text{eV}^2$.}\label{table:parameters}
\end{center}
\end{table*}

\section{Oscillation probabilities and detecting sterile neutrinos} \label{sec:probability}
In this section we will discuss a possible application of the perturbative expressions above for detecting sterile neutrinos.  The principle of the approach is that one can calculate the theoretical predictions of the oscillation probabilities in different schemes and compare them with the experimental results. Usually for a given baseline and neutrino energy the predictions from different schemes are close, therefore it is essential to figure out sufficiently accurate expressions for the oscillation probabilities. A similar discussion can be found in \cite{Fong:2017gke}.

In a scheme with $N$ sterile neutrinos, the neutrino oscillation probabilities for $\nu_{\alpha}\rightarrow\nu_\beta$ ($\alpha,\beta\in\{e,\mu,\tau\}$) are
\begin{align}
P_{\alpha\beta}=\bigg| \sum^{3+N}_{i=1} \textbf{V}^*_{\alpha i}\textbf{V}_{\beta i}\,e^{-i\frac{\lambda^{(ex)}_i\,L}{2E}}\bigg|^2,
\end{align}
where $\lambda^{(ex)}_i$ are exact eigenvalues. We can chose the zeroth order results as an approximation, i.e. we adopt
\begin{equation}
\textbf{V}\simeq\textbf{U}^m_\text{PMNS},
\end{equation}
where $U^m_\text{PMNS}$ is defined in Eq.\,\ref{eq:umpmns} and 
\begin{equation}
\lambda^{(ex)}_i\simeq\lambda_i, \quad i=1,2,3,4
\end{equation}
where $\lambda_{1,2,3}$ are defined in Eq.\,\ref{eq:lambda123} and $\lambda_4=M^2(b)$.
For the mass of the sterile neutrino, since it is significantly larger than the active ones, the oscillations related to it will be averaged out.

Former and running experimental facilities have provided parameter fitting results of neutrino oscillations for different schemes. With these parameters, for future baselines, one can predict the probabilities in different schemes and this is a potential approach to determine the existence of sterile neutrinos \cite{Fong:2017gke}. We present the probabilities given by the 3+1 scheme and the differences of the probabilities $|\big<P_{3+1}\big>-P_{3\nu\text{SM}}|$, in different channels, in Fig.\,\ref{fig:Psm_Psterile_mue}, \ref{fig:Psm_Psterile_mumu} and \ref{fig:Psm_Psterile_mutau}. The probabilities in the Standard Model are given by \cite{Zaglauer:1988gz,Barger:1980tf}; the 3+1 scheme is calculated by the 0th order rotation method developed in this paper. All the parameters are given in Table\,\ref{table:parameters}.

In the figures we can identify several regions in which the differences are significantly larger than errors of the perturbation expansions. For example, in the $\nu_\mu\rightarrow\nu_e$ channel, around the band of $L/E\simeq\,700\,(\text{km/GeV})$, $|\big<P_\text{3+1}\big>-P_{3\nu\text{SM}}|$ may be larger than 0.02, the differences will be even larger than 0.05 if $L/E\gtrsim1500\,\text{km/GeV}$ and the baseline is longer than 500\,km. In this channel baselines of T2K/HyperK, NOVA and DUNE (estimated) are marked \cite{Abe:2016aoo,Acero:2019qcr,Papadimitriou:2018akk}. For the channel of $\nu_\mu\rightarrow\nu_\mu$, shifts from the $3\nu$SM will be more than 0.05 with $L/E\simeq\,1000(\text{km/GeV})$ and the baseline is longer than $1000\,\text{km}$. For the $\nu_\mu\rightarrow\nu_\tau$ channel, the scale of the greatest difference is larger than 0.16 if $L/E\simeq\,500(\text{km/GeV})$ or $\simeq\,1500(\text{km/GeV})$. Future experiments may measure the oscillation probabilities with baselines and neutrino energies in the region of interest predicted above and compare the results with the numerical outcomes.

\begin{figure*}
    \centering
    \includegraphics[width=2.8in]{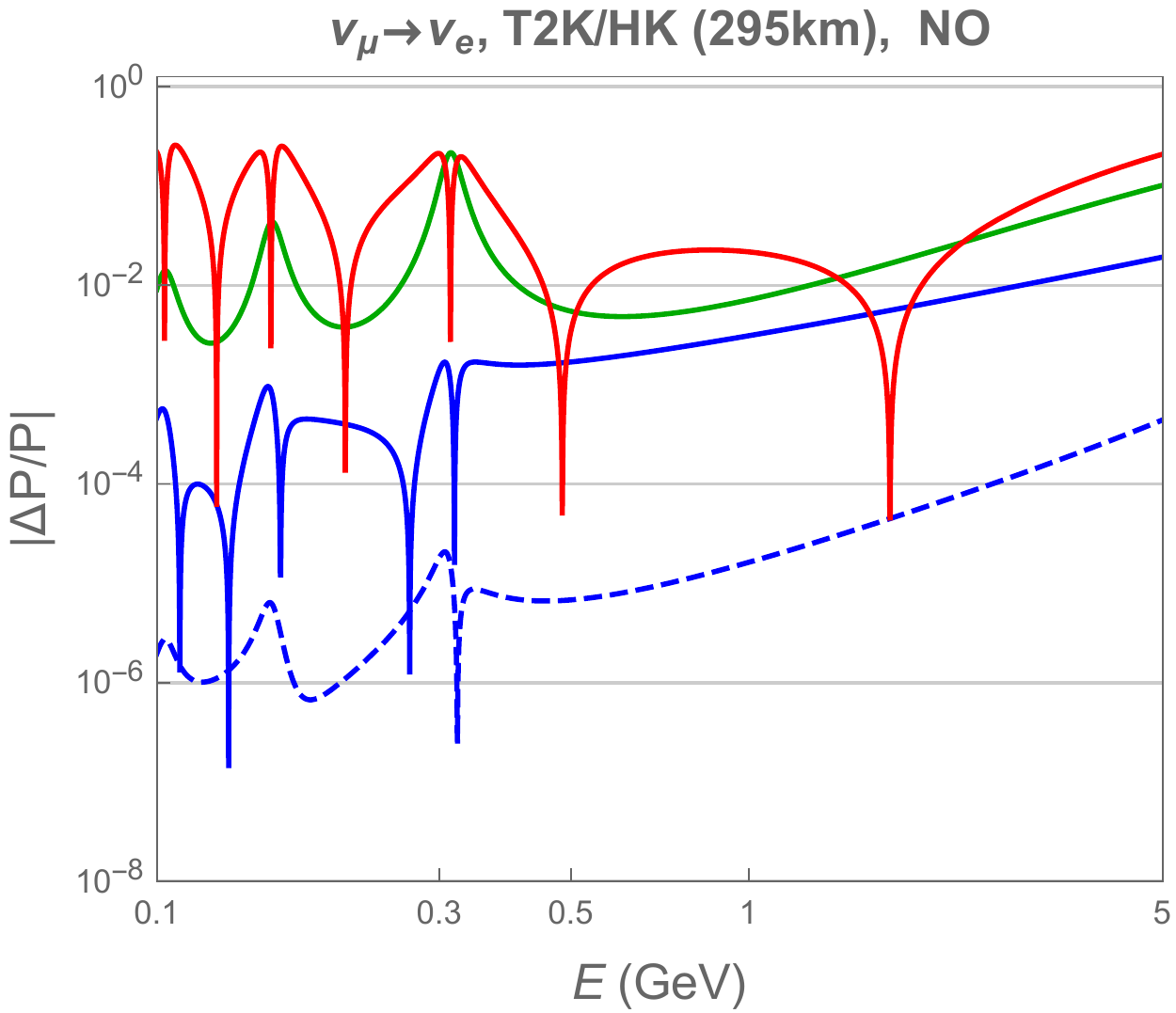}
    \includegraphics[width=2.8in]{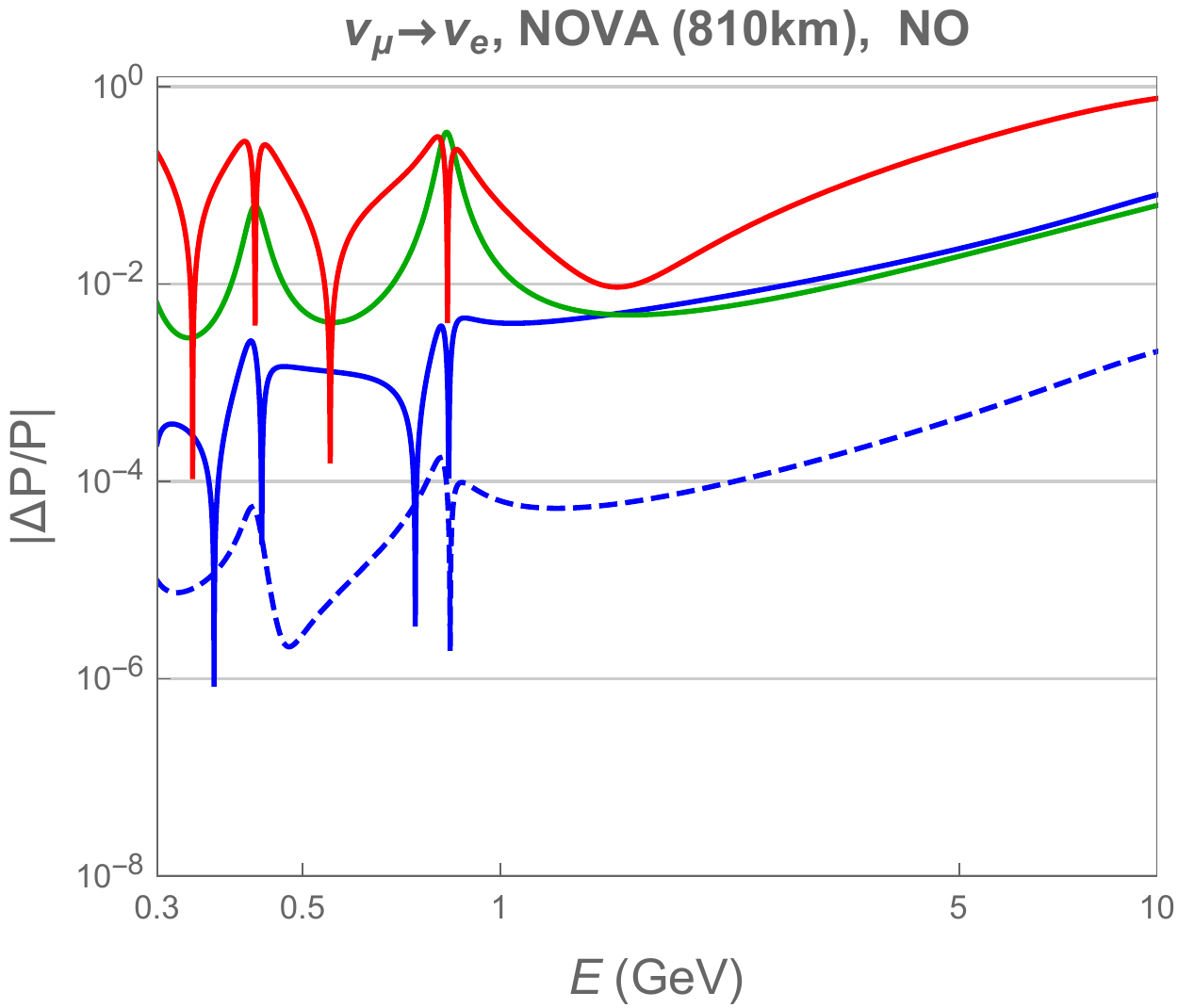}
    \includegraphics[width=2.8in]{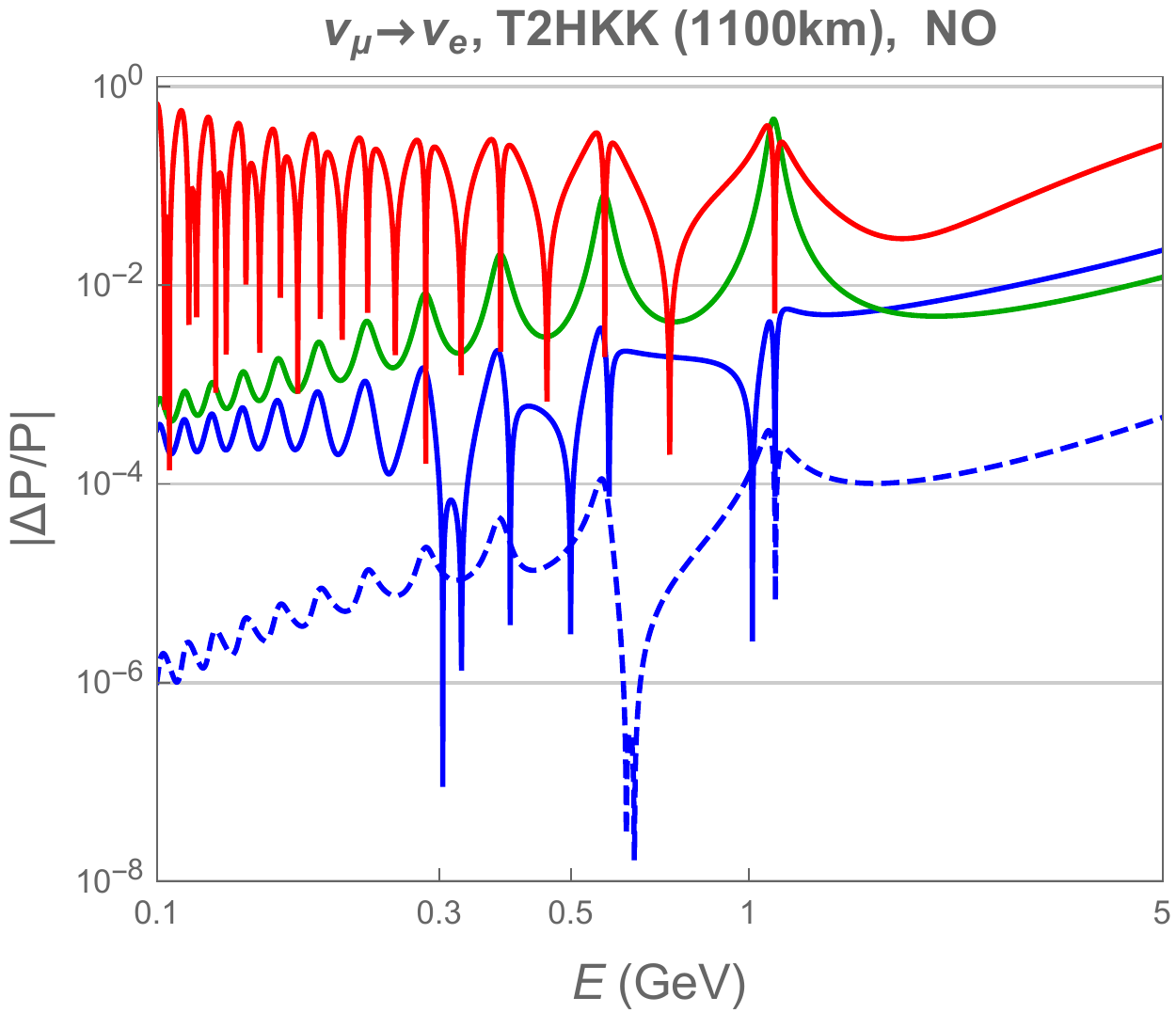}
    \includegraphics[width=2.8in]{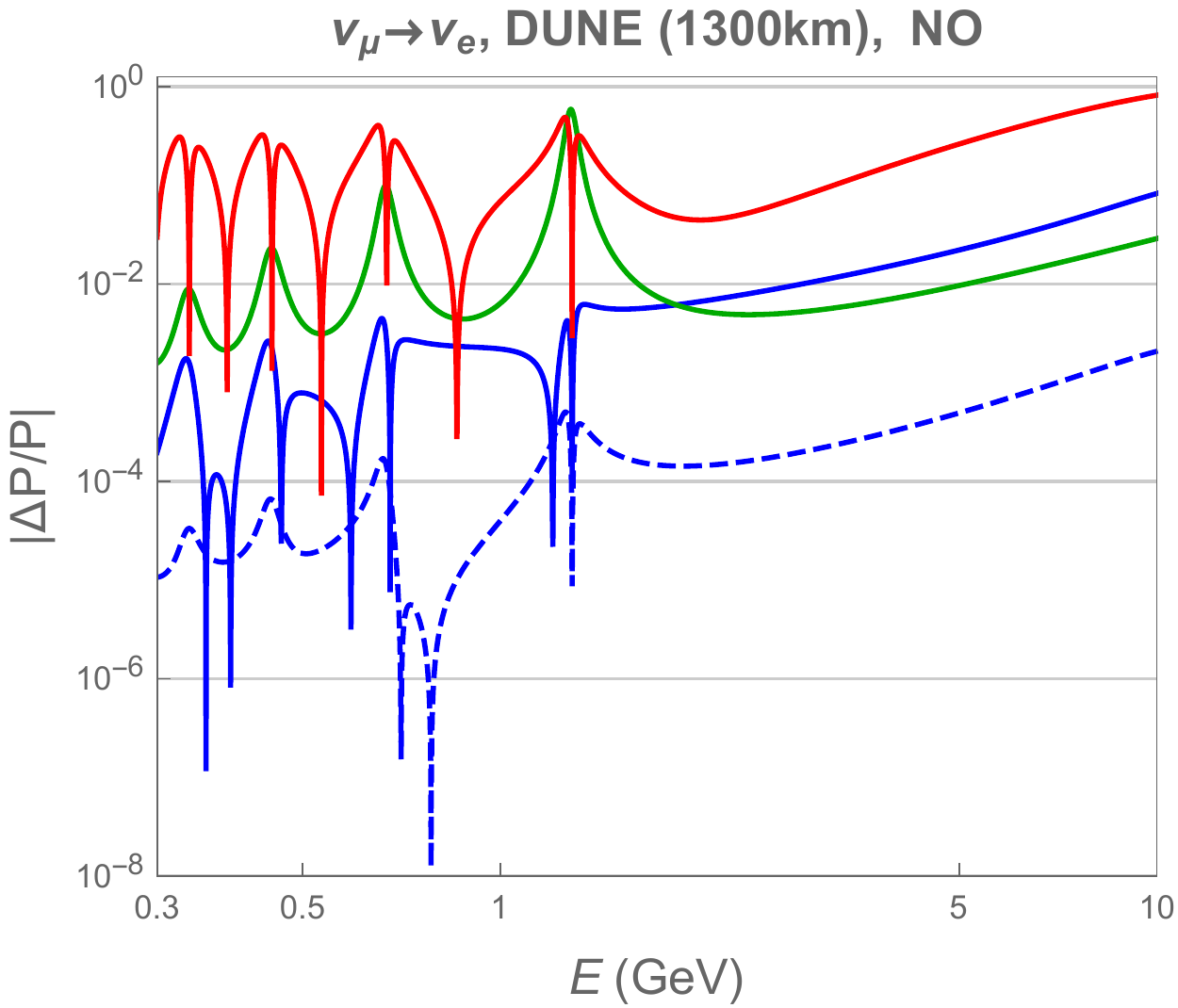}
    \includegraphics[width=4in]{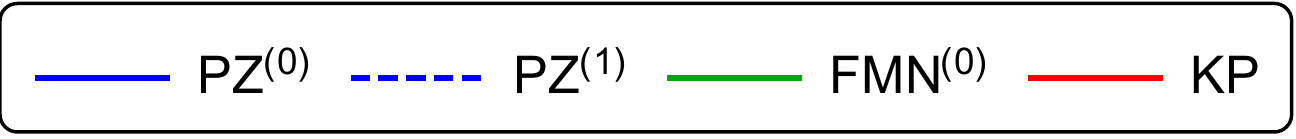}
    \caption{Fractional errors of oscillation probabilities (fast oscillations averaged out) given by different methods. The solid blue curve (PZ) indicates the zeroth order rotation method of this paper and the dashed blue line is the first order result; the green curve (FMN) is from Fong, Minakata and Nunokawa \cite{Fong:2017gke}; the red curve (KP) is from Klop and Palazzo \cite{Klop:2014ima}. Parameters used for this sample calculation are listed in Table\,\ref{table:parameters}. Relative speed of the methods can be found in Table\,\ref{table:computespeed}.
    \label{fig:compare}}
\end{figure*}

\section{Compare to existing approximation formulas}\label{sec:compare}

Approximation methods to calculate neutrino oscillations in matter in 3+1 scheme have been studied by many researchers, for example see \cite{Klop:2014ima,Agarwalla:2016xxa,Agarwalla:2016mrc,Fong:2016yyh,Fong:2017gke}. All these works chose to ignore the Hamiltonian's fourth row/column (except for the fourth diagonal element) in zeroth order approximations; thus the problem was reduced to $3\times 3$ case. However, solving a three dimensional eigensystem is still not simple (many approximation methods developed for the $3\nu SM$ scheme). Fong, Minakata and Numokawa (FMN) adopted the exact 3-dim solutions \cite{Fong:2017gke} which was complicated, see \cite{Zaglauer:1988gz,Kimura:2002wd}. Klop and Palazzo (KP) used one more approximation method for the $3\times 3$ submatrix \cite{Klop:2014ima} which introduced extra errors. Fig.\,\ref{fig:compare} compares fractional errors of the rotations method developed in this paper (PZ) with 0th order FMN and KP, assuming baselines of T2K/Hyper-K (T2K/HK), NOVA, HyperK-Korea (T2HKK) and DUNE. Compared with KP, just to 0th order PZ is significantly more precise for almost all baselines and energy ranges. Based on Fig.\,\ref{fig:compare}, FMN's precision is similar to PZ's for most baselines and energy regions, however, we can still identify PZ's advantage for T2K's baseline or low energy ($\lesssim 1\text{GeV}$) regions.

\begin{table*}[hbt!]
\begin{tabular}{|p{2cm}|p{1.5cm}|p{1.5cm}|p{1.5cm}|p{1cm}|p{2.8cm}|p{2.7cm}|}
\hline
\centering{Method} & \centering{PZ$^{(0)}$} & \centering{PZ$^{(1)}$}&\centering{FMN$^{(0)}$} &\centering{KP}& Analytical Solution& Numerical Method\\
\hline
\centering{Time Units} & \centering{1.0} &\centering{1.8} &\centering{2.1} &\centering{0.16} &\centering{2.2} &\hspace{1.15cm}5.7\\
\hline
\end{tabular}
\caption{Computation time consumed by different methods. Since a real time will depend on a specific computer's performance, 0th order PZ's (rotation method developed in this paper) time is set to be one unit time. FMN is from Fong, Minakata and Nunokawa \cite{Fong:2017gke}; KP is from Klop and Palazzo \cite{Klop:2014ima}; the analytical solution is given by \cite{Li:2018ezt}; the numerical method is referred to Eigen 3.3.7, \href{http://eigen.tuxfamily.org/}{http://eigen.tuxfamily.org}. \label{table:computespeed}}
\end{table*}

We compare computing time of the different methods in Table\,\ref{table:computespeed}, since any specific computing time heavily depends on performance of the computer, we provide a list of relative computing time, i.e 0th order PZ's computing time is set to one unit time. The speed of a numerical method (using Eigen 3.3.7, \href{http://eigen.tuxfamily.org/}{http://eigen.tuxfamily.org}) is also included in the comparison. A similar comparison of the methods for the $3\nu$SM scheme can be found in \cite{Parke:2019vbs}. Table\,\ref{table:computespeed} shows that compared with the rotation method developed in this paper (PZ), only the KP method is faster, however, its advantage in speed will be offset by the relatively poor precision. The FMN method is comparable in time consumed to the exact analytical solution. For experimentalists, the speed of a numerical method for evaluating the oscillation probability is relevant because it effects the time and computing resources consumed by large multi-dimensional parameter scans.  

Besides simplicity and better precision, the rotation method of this paper also gives explicit expressions of zeroth order eigenvalues and mixing angles and CP phases with high precision which are not covered by any former references.

\section{Conclusions}\label{sec:conclusion}
A compact and simple technique for calculating neutrino oscillation probabilities in matter for schemes with sterile neutrino has been developed from the extension of an analogous method for the $3\nu$SM model \cite{Denton:2016wmg}. The extended method is appropriate to conditions in which the Wolfenstein matter potentials defined in Eq.\,\ref{eq:ab} are not significantly larger than $\Delta m^2_\text{atm}$, meaning that it may be applied to all the current and proposed accelerator neutrino oscillation experiments. The zeroth order eigensystem of the Hamiltonian in the active space (i.e. the three dimensions included in $3\nu $SM) derived by the method is non-degenerate. Meanwhile, numerical study shows that crossings of the zeroth order eigenvalues involving the sterile one only happen with large matter potential (high neutrino energy for the earth's crust), which is out of the paper's scope of discussion. An additional crucial advantage of the method developed in this paper is that errors of the zeroth order results are small  when the matter potentials are $\leq \Delta m^2_{ee}$ and vanish in vacuum because the matter potential terms are factors of all the perturbative terms.

We implement a series of complex or real rotations to kill the leading order off-diagonal elements and resolve crossings of the diagonal elements of the Hamiltonian. The rotation angles and phases are the zeroth order mixing parameters of the effective PMNS matrix in matter. In the rotated Hamiltonian, the diagonal elements are the zeroth order eigenvalues, whereas  the off-diagonal elements are the perturbing Hamiltonian. Based on this arrangement, perturbation expansions are performed after the rotations to achieve better accuracy. When the matter effect is comparable to the vacuum mixing effect, i.e. matter potentials defined in Eq.\,\ref{eq:ab} are comparable to $\Delta m^2_{31}$, the expansion parameter is no larger than $\Delta m^2_{21}/\Delta m^2_{31}\simeq 0.03$; when the matter effect itself is weak, the perturbative Hamiltonian will be higher order because it consists of terms proportional to $a\epsilon$ or $b\epsilon$. 

Finally numerical tests show that the rotation method developed in this paper balances precision with computing speed nicely; comparisons with a numerical method and some previous approximate methods have been applied to demonstrate the rotation method's advantages as presented in this paper. For the first order perturbation expansion of this paper, absolute errors of the oscillation probabilities are shown to be no more than $10^{-4}$. This precision is sufficient to distinguish the schemes with sterile neutrinos from the $3\nu$SM model, which makes the method developed by this paper suitable to explore the existence of sterile neutrinos.

\section*{Acknowledgements}
This manuscript has been authored by Fermi Research Alliance, LLC under Contract
No.  DE-AC02-07CH11359 with the U.S. Department of Energy,  Office of Science,
Office of High Energy Physics.

This project has received funding/support from the European Union’s Horizon 2020 research and innovation programme under the Marie Sklodowska-Curie grant agreement No 690575 and No 674896. 

We thank Peter Denton for discussions and comments on this paper and Heath O'Connell for  a careful reading of this manuscript. X.Z. thanks Melissa Clegg for discussing the English expressions in this paper.

\appendix

\section{Mixing angles and phases in the new convention of the PMNS matrix} \label{appendix:convention}
The PMNS matrix in the new and the usual convention can be expressed as
\begin{widetext}
\begin{align}
\textbf{U}^{3+1}_\text{PMNS}\equiv\begin{cases}
 \textbf{U}_{23}(\theta_{23},\delta_{23})\,\textbf{U}_{34}(\theta_{34},\delta_{34})\,\textbf{U}_{24}(\theta_{24},\delta_{24})\,\textbf{U}_{14}(\theta_{14})\,\textbf{U}_{13}(\theta_{13})\,\textbf{U}_{12}(\theta_{12})& \\
 \textbf{U}_{34}(\theta^\prime_{34},\delta^\prime_{34})\,\textbf{U}_{24}(\theta^\prime_{24},\delta^\prime_{24})\,\textbf{U}_{14}(\theta^\prime_{14})\,\textbf{U}_{23}(\theta^\prime_{23},\delta^\prime_{23})\,\textbf{U}_{13}(\theta^\prime_{13})\,\textbf{U}_{12}(\theta^\prime_{12})&
\end{cases}.
\end{align}
\end{widetext}
We will express the parameters of the new convention (without the prime) in formulas of the parameters of the usual convention (with prime). We notice that $\textbf{U}_{14}\textbf{U}_{23}=\textbf{U}_{23}\textbf{U}_{14}$ then 
\begin{equation}
\theta_{12}=\theta^\prime_{12},\quad\theta_{13}=\theta^\prime_{13},\quad\theta_{14}=\theta^\prime_{14},
\end{equation}
and 
\begin{align}
&\textbf{U}_{23}(\theta_{23},\delta_{23})\,\textbf{U}_{34}(\theta_{34},\delta_{34})\,\textbf{U}_{24}(\theta_{24},\delta_{24}) \notag \\
&=e^{i\textbf{A}}\textbf{U}(\theta^\prime_{34},\delta^\prime_{34})\,\textbf{U}_{24}(\theta^\prime_{24},\delta^\prime_{24})\,\textbf{U}_{23}(\theta^\prime_{23},\delta^\prime_{23}), \label{eq:234}
\end{align}
where $\textbf{A}$ is a traceless real diagonal matrix. Solving Eq.\,\ref{eq:234} we get the following relations
\begin{align}
s_{34}&=\big|s^\prime_{34}c^\prime_{23}+s^\prime_{23}s^\prime_{24}c^\prime_{34}\,e^{i(\delta^\prime_{23}-\delta^\prime_{24}+\delta^\prime_{34})}\big|,\notag \\
s_{24}&=\sqrt{1-\Big(\frac{c^\prime_{24}c^\prime_{34}}{c_{34}}\Big)^2},\notag \\
s_{23}&=\frac{s^\prime_{23}c^\prime_{24}}{c_{34}},
\end{align}
and
\begin{align}
\delta_{34}&=\text{Arg}\big[s^\prime_{34}c^\prime_{23}\,e^{i\delta^\prime_{34}}+s^\prime_{23}s^\prime_{24}c^\prime_{34}\,e^{i(\delta^\prime_{24}-\delta^\prime_{23})}\big], \notag \\
\delta_{24}&=\text{Arg}\big[s^\prime_{24}c^\prime_{23}c^\prime_{34}\,e^{i\delta^\prime_{24}}-s^\prime_{23}s^\prime_{34}\,e^{i(\delta^\prime_{23}+\delta^\prime_{34})}\big], \notag \\
\delta_{23}&=\delta^\prime_{23}+\text{Arg}\big[c^\prime_{23}c^\prime_{34}-s^\prime_{23}s^\prime_{24}s^\prime_{34}\,e^{i(\delta^\prime_{23}-\delta^\prime_{24}+\delta^\prime_{34})}\big].
\end{align}
The approximated formulas, with $\mathcal{O}(\epsilon)$ corrections, are also listed below

\begin{align}
s_{34}&\simeq \Big[c^{\prime 2}_{23}s^{\prime 2}_{34}+s^{\prime 2}_{23}c^{\prime 2}_{34}\notag \\
&\qquad +2s^\prime_{23}s^\prime_{24}s^\prime_{34}c^\prime_{23}\cos(\delta^\prime_{23}+\delta^\prime_{34}-\delta^\prime_{24})\Big]^{1/2}+\mathcal{O}(\epsilon), \notag \\
s_{24}&\simeq \Big[c^{\prime 2}_{23}s^{\prime 2}_{24}+s^{\prime 2}_{23}s^{\prime 2}_{34} \notag \\
&\qquad -2s^\prime_{23}s^\prime_{24}s^\prime_{34}c^\prime_{23}\cos(\delta^\prime_{23}+\delta^\prime_{34}-\delta^\prime_{24})\Big]^{1/2}+\mathcal{O}(\epsilon), \notag \\
s_{23}&\simeq s^\prime_{23}+\mathcal{O}(\epsilon), \notag \\
\delta_{34}&\simeq \arctan\frac{s^\prime_{23}s^\prime_{24}\sin(\delta^\prime_{24}-\delta^\prime_{23})+s^\prime_{34}c^\prime_{23}\sin\delta^\prime_{34}}{s^\prime_{23}s^\prime_{24}\cos(\delta^\prime_{24}-\delta^\prime_{23})+s^\prime_{34}c^\prime_{23}\cos\delta^\prime_{34}} \notag \\
&\qquad+\frac{\pi}{2}\big[1-\text{sign}(s^\prime_{34}c^\prime_{23}+s^\prime_{23}s^\prime_{24}c^\prime_{34})\big]+\mathcal{O}(\epsilon), \notag \\
\delta_{24}&\simeq \arctan\frac{c^\prime_{23}s^\prime_{24}\sin\delta^\prime_{24}-s^\prime_{23}s^\prime_{34}\sin(\delta^\prime_{23}+\delta^\prime_{34})}{c^\prime_{23}s^\prime_{24}\cos\delta^\prime_{24}-s^\prime_{23}s^\prime_{34}\cos(\delta^\prime_{23}+\delta^\prime_{34})} \notag \\
&\qquad+\frac{\pi}{2}\big[1-\text{sign}(s^\prime_{24}c^\prime_{23}c^\prime_{34}-s^\prime_{23}s^\prime_{34})\big]+\mathcal{O}(\epsilon), \notag \\
\delta_{23}&\simeq \delta^\prime_{23}+\mathcal{O}(\epsilon).
\end{align}

\section{Complex phases convention} \label{appendix:phase}
In Section \ref{subsec:convention} we chose $\textbf{U}_{12}$ and $\textbf{U}_{13}$ to be real; however, now $\alpha_{12}$ and $\alpha_{13}$ are non-zero. To recover the initial convention of the complex phases we need to implement a phase transformation. Firstly, we multiply the 1st row by $e^{-i\,\alpha_{12}}$ and the 1st column by $e^{i\,\alpha_{12}}$; then the 3rd row is multiplied by $e^{i(\alpha_{13}-\alpha_{12})}$ and the 3rd column is multiplied by $e^{i(-\alpha_{13}+\alpha_{12})}$. Finally all the complex phases are absorbed into $\textbf{U}_{23}$, $\textbf{U}_{24}$ and $\textbf{U}_{34}$. The zeroth order phases are
\begin{align}
\tilde{\delta}_{12}&=0, \notag \\
\tilde{\delta}_{13}&=0, \notag \\
\tilde{\delta}_{23}&=\delta_{23}-\alpha_{13}+\alpha_{12}, \notag \\
\tilde{\delta}_{24}&=\delta_{24}+\alpha_{12}, \notag \\
\tilde{\delta}_{34}&=\delta_{34}+\alpha_{13}.
\end{align}

\section{Elements of $\check{\textbf{H}}_M$}\label{appendix:checkHM}
Since the Hamiltonian must be Hermitian, we will just present the 4th column. 

\begin{align}
(\check{\textbf{H}}_M)_{14}=& \frac{1}{2E}\Big[\tilde{c}_{12}\tilde{c}_{13}s_{14}c_{14}\big(a+b\,c^2_{24}c^2_{34}\big)\notag\\
  &\qquad-b\,\tilde{s}_{12}s_{24}c_{14}c_{24}c^2_{34}\,e^{i(\delta_{24}+\alpha_{12})} \notag \\
  &\qquad-b\,\tilde{s}_{13}\tilde{c}_{12}s_{34}c_{14}c_{24}c_{34}\,e^{i(\delta_{34}+\alpha_{13})}\Big],\notag \\
(\check{\textbf{H}}_M)_{24}=&\frac{1}{2E}\Big[\tilde{s}_{12}\tilde{c}_{13}s_{14}c_{14}\big(a+b\,c^2_{24}c^2_{34}\big)\notag\\
&\qquad+b\,\tilde{c}_{12}s_{24}c_{14}c_{24}c^2_{34}\,e^{i\delta_{24}}\notag \\
&\qquad-b\,\tilde{s}_{12}\tilde{s}_{13}s_{34}c_{14}c_{24}c_{34}\,e^{i(\delta_{34}-\alpha_{12}+\alpha_{13})}\Big], \notag \\
(\check{\textbf{H}}_M)_{34}=&\frac{1}{2E}\Big[\tilde{s}_{13}s_{14}c_{14}\big(a+b\,c^2_{24}c^2_{34}\big)e^{-i\alpha_{13}}\notag \\
&\qquad+b\,\tilde{c}_{13}s_{34}c_{14}c_{24}c_{34}\,e^{i\delta_{34}}\Big].
\end{align}

\section{Perturbation expansions}\label{appendix:perturbation}

\subsection{First order corrections}

Since all diagonal elements have been absorbed into the zeroth order Hamiltonian, by Eq.\,\ref{eq:lambdai1}, the first order corrections to the eigenvalues, which are the diagonal elements of the perturbative Hamiltonian, are zero, i.e.
\begin{align}
\lambda^{(1)}_i=2E\,(\check{\textbf{H}}_1)_{ii}=0.
\end{align}

As for the eigenvectors, to first order, are derived from Eq.\,\ref{eq:w1} as follows:
\begin{widetext}
\begin{equation}
\textbf{W}_1=\frac{\epsilon^\prime\Delta m^2_{ee}}{2E}\left(\begin{array}{cccc}
 0&0 &-\frac{\tilde{s}_{12}}{\Delta \lambda_{31}}e^{i(\alpha_{12}+\alpha_\epsilon)} &0 \\
 0&0 &\frac{\tilde{c}_{12}}{\Delta\lambda_{32}}e^{i\alpha_\epsilon} &0 \\
\frac{\tilde{s}_{12}}{\Delta \lambda_{31}}e^{-i(\alpha_{12}+\alpha_\epsilon)} &-\frac{\tilde{c}_{12}}{\Delta\lambda_{32}}e^{-i\alpha_\epsilon} & 0&0\\
0& 0&0 & 0
\end{array}\right)+(2E)\left(\begin{array}{cccc}
0&0 &0 &-\frac{(\check{\textbf{H}}_M)_{14}}{\lambda_1-M^2} \\
0&0 &0 &-\frac{(\check{\textbf{H}}_M)_{24}}{\lambda_2-M^2} \\
0&0 &0 &-\frac{(\check{\textbf{H}}_M)_{34}}{\lambda_3-M^2} \\
\frac{(\check{\textbf{H}}_M)^*_{14}}{\lambda_1-M^2} & \frac{(\check{\textbf{H}}_M)^*_{24}}{\lambda_2-M^2} & \frac{(\check{\textbf{H}}_M)^*_{34}}{\lambda_3-M^2} &0
\end{array}\right).
\end{equation}
\end{widetext}

\subsection{Second order corrections}
The second order corrections to eigenvalues are
\begin{equation}
\lambda_i^{(2)}=\sum_{i\neq k}\frac{|2E(\check{\textbf{H}}_1)_{ik}|^2}{\lambda_i-\lambda_k}, \label{eq:lambdai2}
\end{equation}
or explicitly as
\begin{align}
\lambda^{(2)}_1=&-(\epsilon^\prime\Delta m^2_{ee})^2\,\frac{\tilde{s}^2_{12}}{\Delta\lambda_{31}}+\frac{|2E(\check{\textbf{H}}_M)_{14}|^2}{\lambda_1-M^2},\notag \\
\lambda^{(2)}_2=&-(\epsilon^\prime\Delta m^2_{ee})^2\,\frac{\tilde{c}^2_{12}}{\Delta\lambda_{32}}+\frac{|2E(\check{\textbf{H}}_M)_{24}|^2}{\lambda_2-M^2}, \notag \\
\lambda^{(2)}_3=&(\epsilon^\prime\Delta m^2_{ee})^2\,\big(\frac{\tilde{s}^2_{12}}{\Delta\lambda_{31}}+\frac{\tilde{c}^2_{12}}{\Delta\lambda_{32}}\big)+\frac{|2E(\check{\textbf{H}}_M)_{34}|^2}{\lambda_3-M^2}, \notag \\
\lambda_4^{(2)}=&-\frac{|2E(\check{\textbf{H}}_M)_{14}|^2}{\lambda_1-M^2}-\frac{|2E(\check{\textbf{H}}_M)_{24}|^2}{\lambda_2-M^2}-\frac{|2E(\check{\textbf{H}}_M)_{34}|^2}{\lambda_3-M^2}. \label{eq:lambda2order}
\end{align}
where $\Delta \lambda_{ij}\equiv \lambda_i-\lambda_j$. 

The second corrections to eigenvectors can be calculated by the corrections to the PMNS matrix:
\begin{equation}
(\textbf{W}_2)_{ij}=\begin{cases}
-\frac{1}{2}\sum\limits_{k\neq i}\frac{|2E(\check{\textbf{H}}_1)_{ik}|^2}{(\lambda_i-\lambda_k)^2} & i=j \\
\frac{1}{\lambda_i-\lambda_j}\sum\limits_{k\neq i, k\neq j}\frac{2E(\check{\textbf{H}}_1)_{ik}2E(\check{\textbf{H}}_1)_{kj}}{\lambda_k-\lambda_j} & i\neq j
\end{cases}.\label{eq:w2}
\end{equation}
We list the elements of $\textbf{W}_2$ below:
\begin{align}
(\textbf{W}_2)_{11}&=-(\epsilon^\prime\Delta m^2_{ee})^2\frac{\tilde{s}^2_{12}}{2(\Delta \lambda_{31})^2} -\frac{(2E)^2|(\check{\textbf{H}}_M)_{14}|^2}{2(M^2-\lambda_1)^2}, \notag \\
(\textbf{W}_2)_{12}&=(\epsilon^\prime\Delta m^2_{ee})^2\frac{\tilde{s}_{12}\tilde{c}_{12}\,e^{i\alpha_{12}}}{\Delta\lambda_{32}\Delta\lambda_{21}}\notag \\
&\qquad-(2E)^2\frac{(\check{\textbf{H}}_M)_{14}(\check{\textbf{H}}_M)^*_{24}}{(M^2-\lambda_2)\Delta\lambda_{21}}, \notag \\
(\textbf{W}_2)_{13}&=-(2E)^2\frac{(\check{\textbf{H}}_M)_{14}(\check{\textbf{H}}_M)^*_{34}}{(M^2-\lambda_3)\Delta\lambda_{31}}, \notag \\
(\textbf{W}_2)_{21}&=-(\epsilon^\prime\Delta m^2_{ee})^2\frac{\tilde{s}_{12}\tilde{c}_{12}\,e^{-i\alpha_{12}}}{\Delta\lambda_{31}\Delta\lambda_{21}}\notag \\
&\hspace*{2cm}+(2E)^2\frac{(\check{\textbf{H}}_M)_{24}(\check{\textbf{H}}_M)^*_{14}}{(M^2-\lambda_1)\Delta\lambda_{21}}, \notag \\
(\textbf{W}_2)_{22}&=-(\epsilon^\prime\Delta m^2_{ee})^2\frac{\tilde{c}^2_{12}}{2(\Delta\lambda_{32})^2}-(2E)^2\frac{|(\check{\textbf{H}}_M)_{24}|^2}{2(M^2-\lambda_2)^2}, \notag \\
(\textbf{W}_2)_{23}&=-(2E)^2\frac{(\check{\textbf{H}}_M)_{24}(\check{\textbf{H}}_M)^*_{34}}{(M^2-\lambda_3)\Delta\lambda_{32}}, \notag \\
(\textbf{W}_2)_{31}&=(2E)^2\frac{(\check{\textbf{H}}_M)_{34}(\check{\textbf{H}}_M)^*_{14}}{(M^2-\lambda_1)\Delta\lambda_{31}}, \notag \\
(\textbf{W}_2)_{32}&=(2E)^2\frac{(\check{\textbf{H}}_M)_{34}(\check{\textbf{H}}_M)^*_{24}}{(M^2-\lambda_2)\Delta\lambda_{32}}, \notag \\
(\textbf{W}_2)_{33}&=-\frac{(\epsilon^\prime\Delta m^2_{ee})^2}{2}\Big[\frac{\tilde{s}^2_{12}}{(\Delta\lambda_{31})^2}+\frac{\tilde{c}^2_{12}}{(\Delta\lambda_{32})^2}\Big]\notag \\
&\hspace*{2cm}-(2E)^2\frac{|(\check{\textbf{H}}_M)_{34}|^2}{2(M^2-\lambda_3)^2}, \notag \\
(\textbf{W}_2)_{14}&=-\epsilon^\prime\Delta m^2_{ee}\frac{(2E)\tilde{s}_{13}(\check{\textbf{H}}_M)_{34}e^{i(\alpha_{13}+\alpha_\epsilon)}}{(\lambda_2-M^2)(\lambda_3-M^2)}, \notag \\
(\textbf{W}_2)_{24}&= \epsilon^\prime\Delta m^2_{ee}\frac{(2E)\tilde{c}_{13}(\check{\textbf{H}}_M)_{34}e^{i\alpha_\epsilon}}{(\lambda_1-M^2)(\lambda_3-M^2)}, \notag \\
(\textbf{W}_2)_{34}&=-\epsilon^\prime\Delta m^2_{ee}(2E)e^{-i\alpha_\epsilon}\bigg[\frac{\tilde{s}_{13}(\check{\textbf{H}}_M)_{14}e^{-i\alpha_{13}}}{(\lambda_3-M^2)(\lambda_1-M^2)}\notag \\
&\hspace*{3.5cm}+\frac{\tilde{c}_{13}(\check{\textbf{H}}_M)_{24}}{(\lambda_3-M^2)(\lambda_2-M^2)} \bigg],\notag \\
(\textbf{W}_2)_{41}&=-\epsilon^\prime\Delta m^2_{ee}\frac{(2E)\tilde{s}_{13}(\check{\textbf{H}}_M)^*_{34}e^{-i(\alpha_{13}+\alpha_\epsilon)}}{(M^2-\lambda_1)\Delta\lambda_{31}}, \notag \\
(\textbf{W}_2)_{42}&=\epsilon^\prime\Delta m^2_{ee}\frac{(2E)\tilde{c}_{13}(\check{\textbf{H}}_M)^*_{34}e^{-i\alpha_\epsilon}}{(M^2-\lambda_2)\Delta\lambda_{32}}, \notag \\
(\textbf{W}_2)_{43}&=\epsilon^\prime\Delta m^2_{ee}(2E)e^{i\alpha_\epsilon}\bigg[\frac{\tilde{s}_{13}(\check{\textbf{H}}_M)^*_{34}e^{i\alpha_{13}}}{(M^2-\lambda_3)\Delta\lambda_{31}}\notag \\
&\hspace*{3.5cm}-\frac{\tilde{c}_{13}(\check{\textbf{H}}_M)^*_{34}}{(M^2-\lambda_3)\Delta\lambda_{32}}\bigg], \notag \\
(\textbf{W}_2)_{44}&=-2E^2\Big[\frac{|(\check{\textbf{H}}_M)_{14}|^2}{(M^2-\lambda_1)^2}\notag \\
&\hspace*{2cm}+\frac{|(\check{\textbf{H}}_M)_{24}|^2}{(M^2-\lambda_2)^2}+\frac{|(\check{\textbf{H}}_M)_{34}|^2}{(M^2-\lambda_3)^2}\Big]. \\
\notag
\end{align}

\bibliography{main}

\end{document}